\documentclass[prd,onecolumn,amsmath,amssymb,floatfix,superscriptaddress,notitlepage,nofootinbib,preprintnumbers]{revtex4-1}
\usepackage{amsfonts,amssymb,amsmath,graphicx,color,bm}
\definecolor{ultramarine}{rgb}{0.07, 0.04, 0.56}
\definecolor{cadmiumgreen}{rgb}{0.0, 0.42, 0.24}
\definecolor{indigo(dye)}{rgb}{0.0, 0.25, 0.42}
\usepackage[linktocpage=true]{hyperref}
\hypersetup{
colorlinks=true,
citecolor=ultramarine,
linkcolor=cadmiumgreen,
urlcolor=indigo(dye),
}

\newcommand{\f}[2]{\frac{#1}{#2}}  
\newcommand{\mk}[1]{\left( #1 \right)}  
\newcommand{\kk}[1]{\left[ #1 \right]}  
\newcommand{\ck}[1]{\left\{ #1 \right\}}  
\newcommand{\be}{\begin{equation}}  
\newcommand{\ee}{\end{equation}}
\newcommand{\bem}{\begin{pmatrix}}
\newcommand{\eem}{\end{pmatrix}}
\newcommand{\Mpl}{M_{\rm Pl}}
\renewcommand{\d}{\delta}
\newcommand{\e}{\epsilon}

\newcommand{\C}{\mathcal{C}}
\newcommand{\F}{\mathcal{F}}

\renewcommand{\L}{\mathcal{L}}

\newcommand{\mS}{K^i_{\hphantom{i}j} K^j_{\hphantom{i}i}}
\newcommand{\T}{K^i_{\hphantom{i}j} K^j_{\hphantom{i}k} K^k_{\hphantom{i}i}}
\newcommand{\V}{\alpha^i_{\hphantom{i}j} K^j_{\hphantom{i}i}}
\newcommand{\al}{\alpha}

\renewcommand{\O}{\mathcal{O}}

\newcommand{\pa}{\partial}

\newcommand{\N}{\mathcal{N}}

\newcommand{\comment}[1]{\textcolor{green}{}}


\begin{document}%%%%%%%%%%%%%%%%%%%%%%%%%%%%%%%%%%%%%%%%%  

\title{Effective field theory of degenerate higher-order inflation}

\author{Hayato Motohashi$^1$\footnote{Present address: Division of Liberal Arts, Kogakuin University, 2665-1 Nakano-machi, Hachioji, Tokyo, 192-0015, Japan}}
\noaffiliation
\affiliation{Center for Gravitational Physics, Yukawa Institute for Theoretical Physics, Kyoto University, Kyoto 606-8502, Japan}

\author{Wayne Hu}
\affiliation{Kavli Institute for Cosmological Physics, Department of Astronomy and Astrophysics, Enrico Fermi Institute, The University of Chicago, Chicago, Illinois 60637, USA}

\preprint{YITP-20-09}

\begin{abstract}%%%%%%%%%%%%%%%%%%%%%%%%%%%%%%%%%%%%%%%%%  
We extend the effective field theory of inflation to a general Lagrangian constructed from Arnowitt-Deser-Misner variables that encompasses the most general interactions with up to second derivatives of the scalar field whose background breaks temporal diffeomorphism invariance.
Degeneracy conditions, corresponding to  8 distinct types -- only one of which corresponds to known degenerate higher-order scalar-tensor models -- provide necessary conditions for eliminating
the Ostrogradsky ghost in a  covariant theory at the level of the quadratic action in
unitary gauge.  Novel implications of the degenerate higher-order system 
for the Cauchy problem are illustrated with the phase space portrait of an explicit inflationary example:  not all field configurations lead 
to physical solutions for the metric even for positive potentials; solutions are unique for a given configuration only up to a branch choice;
solutions  on one branch can apparently end at nonsingular points of the metric and their continuation on
alternate branches lead to nonsingular bouncing solutions;  unitary gauge perturbations can go unstable even when 
degenerate terms in the Lagrangian are infinitesimal.   The attractor solution leads to an inflationary scenario where
slow-roll parameters  vary and running of the tilt  can be large even with no explicit features in the potential far from the end of inflation, requiring the optimized slow-roll approach for predicting observables.
\end{abstract}

% 98.80.-k Cosmology
% 98.80.Cq Particle-theory and field-theory models of the early Universe (including cosmic pancakes, cosmic strings, chaotic phenomena, inflationary universe, etc.)  

\date{\today}

\maketitle 

%%%%%%%%%%%%%%%%%%%%%%%%%%%%%%%%%%%%%%%%%  

\section{Introduction}%%%%%%%%%%%%%%%%%%%%%%%%%%%%%%%%%%%%%%%%%
\label{sec:int}

Single-field scalar-tensor theories as inflationary models can be studied in a unified way in the framework of the effective field theory (EFT) of inflation, where the timelike scalar field is treated as a clock that breaks the time diffeomorphism invariance leaving spatial diffeomorphism invariance unbroken~\cite{Creminelli:2006xe,Cheung:2007st}.
In general, the EFT of inflation with higher-derivative operators contains extra ghost degrees of freedom, which may or may not propagate in the regime of validity of the EFT.
To consider a regime where higher-derivative interactions also produce interesting
observable phenomenology, one needs to rely on the framework where the ghost degrees of freedom are appropriately eliminated. 
Therefore, an EFT Lagrangian motivated by general ghost-free theories serve as a useful framework. 
In this context, the original EFT framework has been extended in subsequent works~\cite{Gleyzes:2013ooa,Kase:2014cwa,Gleyzes:2014rba,Gleyzes:2015pma,Motohashi:2017gqb}
to include derivative operators appearing in the Horndeski \cite{Horndeski:1974wa,Nicolis:2008in,Deffayet:2009wt,Deffayet:2009mn,Deffayet:2011gz,Kobayashi:2011nu},
Gleyzes-Langlois-Piazza-Vernizzi (GLPV) \cite{Gleyzes:2014dya,Gleyzes:2014qga} 
and Horava-Lifshitz \cite{Horava:2009uw,Blas:2009qj,Blas:2009ck} theories.

More general theories involving additional derivative of fields typically propagate ghost degrees of freedom.
The ghosts associated with higher-order derivatives are known as the Ostrogradsky ghosts~\cite{Ostrogradsky:1850fid,Woodard:2015zca}, which makes the Hamiltonian unbounded due to its linear dependence on canonical momenta. 
Unlike the classically unbounded Hamiltonian of the hydrogen atom, the Ostrogradsky Hamiltonian remains unbounded quantum mechanically as well~\cite{Raidal:2016wop,Smilga:2017arl,Motohashi:2020psc}.
To eliminate the ghost degrees of freedom, one needs to evade the condition of the Ostrogradsky theorem that the Lagrangian is nondegenerate with respect to the highest-order derivatives. 
However,  degeneracy with respect to the highest-order derivative is necessary but not sufficient to evade the unbounded Hamiltonian~\cite{Motohashi:2014opa}, which is the reason why one needs to impose a certain set of degeneracy conditions to eliminate all aspects of the Ostrogradsky ghosts~\cite{Langlois:2015cwa,Motohashi:2016ftl,Motohashi:2017eya,Motohashi:2018pxg}.
This argument can be also understood in a broader context in the language of constraints as a generalization of the Ostrogradsky theorem~\cite{Aoki:2020gfv}.

The degeneracy conditions were applied to a construction of degenerate higher-order scalar-tensor (DHOST) theories with quadratic~\cite{Langlois:2015cwa} and cubic interactions~\cite{BenAchour:2016fzp} of second derivatives of a scalar field, which include the derivative of the lapse function.  The EFT  description of the quadratic and cubic DHOST theories was developed in \cite{Langlois:2017mxy}, where the quadratic Lagrangian around the cosmological background was investigated.
Cosmological evolution and the linear stability analysis were also investigated~\cite{Crisostomi:2018bsp}, focusing on the de Sitter attractor
in a shift symmetric quadratic DHOST model.
In \cite{Langlois:2017mxy,Crisostomi:2018bsp}, two assumptions on background dynamics were adopted: that the lapse remains unity and that the scalar field is proportional to time coordinate. 
The first can be imposed as a gauge condition, and the second  should be satisfied dynamically.
For general timelike scalar field evolution on a given phase space trajectory, one needs to perform a redefinition of the scalar field to make it proportional to time, which changes all of the DHOST coefficients (cf. \cite{Crisostomi:2018bsp} v2).
A dynamical lapse is also taken into account in \cite{Gao:2018znj,Gao:2019lpz} in the context of spatially covariant gravity~\cite{Gao:2014soa},
where the degeneracy conditions were also studied.
However, the general EFT framework of degenerate theories including but not limited to quadratic and cubic DHOST and its application to cosmology have not been fully investigated yet.

In this paper, generalizing our previous work~\cite{Motohashi:2017gqb}, we develop the EFT framework of general degenerate theories, and explore its peculiar phenomenology.  
This framework includes quadratic and cubic DHOST as a special subclass as well as theories where the lapse is nondynamical, e.g.~those with second-order equations of motion for the scalar field, as in the Horndeski case, or the spatial metric in unitary gauge, as in the GLPV case. 
In \S\ref{sec:lag}, we consider the EFT action composed of Arnowitt-Deser-Misner (ADM) geometric quantities including the acceleration and lapse derivative and their couplings to intrinsic and extrinsic curvatures.  
This action includes operators appearing in  covariant theories involving the most general combination of second-order derivatives of scalar field.  
It also includes Lorentz-violating theories such as Horava-Lifshitz gravity~\cite{Horava:2009uw,Blas:2009qj,Blas:2009ck}, as well as the scordatura degenerate theory~\cite{Motohashi:2019ymr} weakly violating the degeneracy condition.
We derive the background and quadratic actions for
various degeneracy classes, which include known DHOST cases,  summarized in Appendix~\ref{app:deg}.   
In \S\ref{sec:inf}, we investigate dynamics in degenerate higher-order inflation, which we dub ``D-inflation'', and clarify several novel features of degenerate models  for both the background and perturbations.  We provide a detailed study of a specific model, for which the optimized slow-roll (OSR) formalism~\cite{Motohashi:2015hpa} serves as a powerful tool as the EFT coefficients can exhibit variation on the several efold time scale.
In \S\ref{sec:con}, we discuss conclusions.

\section{EFT of inflation}%%%%%%%%%%%%%%%%%%%%%%%%%%%%%%%%%%%%%%%%%
\label{sec:lag}

In this section we adopt ADM decomposition and consider the general EFT Lagrangian allowing the most general combination of second-order derivatives of scalar field.
In \S\ref{sec:lag-1}, we construct the EFT Lagrangian from geometric quantities including the acceleration and lapse derivative and their arbitrary couplings to intrinsic and extrinsic curvatures.
In \S\ref{sec:lag-2}, we write down the background and quadratic Lagrangians around cosmological background.
Since vector and tensor perturbations are the same as the previous work~\cite{Motohashi:2017gqb}, in \S\ref{sec:scalarperts} we focus on the scalar perturbations, and reduce the quadratic Lagrangian for a specific degeneracy class.
We provide the complete analysis of the construction of degeneracy conditions in Appendix~\ref{app:deg}.

\subsection{ADM EFT Lagrangian}%%%%%%%%%%%%%%%%%%%%%
\label{sec:lag-1}

We work in the $3+1$ ADM decomposition of the metric
\be ds^2 = -N^2 dt^2 + h_{ij} (dx^i+N^idt)(dx^j+N^jdt) , \ee
where $N$, $N^i$,  $h_{ij}$ are the lapse, shift, spatial metric, respectively.
We define a timelike unit vector $n_\mu\equiv -Nt_{,\mu}$ orthogonal to constant $t$ surfaces,
the acceleration $a_\mu \equiv n^\nu n_{\mu;\nu}$, 
and the extrinsic curvature $K_{\mu\nu}=n_{\nu;\mu}+n_\mu a_\nu$, 
where semicolons on indices here and throughout denote
covariant derivatives with respect to $g_{\mu\nu}$.

For general scalar-tensor theories, we can choose so-called unitary gauge, where $\phi=\phi(t)$ as long
as the gradient of the scalar field is always timelike. 
In  unitary gauge, any Lagrangian with up to second derivatives in the field can be expressed in terms of ADM quantities through
\begin{align} \label{fmn}
\phi_{,\mu} &= - \sqrt{-X} n_\mu ,\notag\\ 
\phi_{;\mu\nu} &= \sqrt{-X} ( - K_{\nu\mu} + n_\nu a_\mu + n_\mu a_\nu - \beta n_\mu n_\nu ) .
\end{align} 
Here, we define $\beta$ by
\begin{align} \label{defbeta}
\beta = -\frac{1}{2}{n^\mu (\ln X)_{,\mu}} {= -\f{\ddot\phi}{N\dot\phi} + \f{\dot N-N^i \pa_i N}{N^2} },
\end{align}
where $X\equiv g^{\mu\nu} \phi_{,\mu}\phi_{,\nu}= -\dot\phi^2/N^2$ is the kinetic term for the scalar.
In particular, if we take $\phi=t$,  $X=-1/N^2$ and 
then $\beta$ measures the fractional change in the lapse along the normal 
\begin{align} 
\beta = n^\lambda (\ln N)_{,\lambda} = \f{\dot N-N^i \pa_i N}{N^2} ,
\end{align}
and more generally it determines the fractional change in the elapsed proper time in field coordinates and so \eqref{defbeta} involves $\ddot \phi$.
In the gauge where $\phi \propto t$, $\ddot \phi=0$  and this has been used widely for the purpose of counting the number of degrees of freedom through the Hamiltonian analysis.  However, to keep a normal perturbation analysis where the background lapse $\bar N=1$, we use $\phi=\phi(t)$, and retain the $\ddot\phi$ term. 
After solving for a given trajectory $\phi(t)$, we can always make a field redefinition $\varphi \propto t(\phi)$, which maintains $\bar N=1$ at the expense of redefining the scalar field Lagrangian, but adopting this at the outset  prevents
a phase space analysis for background trajectories.

We seek to construct a general spatial diffeomorphism invariant EFT Lagrangian involving no more than second-order derivatives of the scalar field which  a priori contains $N, K_{\mu\nu}, a_\mu, \beta$.
So long as we consider theories involving up to $\phi_{;\mu\nu}$, $a^i$ and $a_j$ are the only quantities that have one spatial sub/superscript.
Hence, $a^i$ and $a_j$ always appear together through
\begin{equation}
{\alpha^i}_j \equiv a^i a_j = h^{ik} (\ln N)_{,j} (\ln N)_{,k}, \quad \alpha\equiv {\alpha^i}_i.
\end{equation}
We therefore consider the Lagrangian to be a spatially diffeomorphism invariant function of these
quantities
\be \label{LagEFT} S = \int d^4x N \sqrt{h} L (N, {K^i}_j, {R^i}_j, {\alpha^i}_j, \beta ;t) ,  \ee
generalizing \cite{Motohashi:2017gqb} to allow it to depend on ${\alpha^i}_j$ and $\beta$. 
Here ${R^i}_j$ is the Ricci 3-tensor on the spatial slice. 
Higher-derivative Lagrangians typically contain Ostrogradsky ghosts, but we shall see in the next section that for special combinations of
$N, {K^i}_j, {R^i}_j, {\alpha^i}_j, \beta$ the Lagrangian only propagates one scalar and the usual two tensor
degrees of freedom.
Without the ${\alpha^i}_j$ dependence and allowing higher-order spatial derivatives, the Lagrangian~\eqref{LagEFT} reduces to the one explored in \cite{Gao:2018znj,Gao:2019lpz}.

To explicitly relate this EFT Lagrangian to known ghost-free scalar-tensor theories, we begin with
the most general covariant Lagrangian that is at most cubic in second derivatives
of the field and coupled to the metric as
\be \label{fullDHOST} S = \int d^4x \sqrt{-g} \kk{ F_0+F_1\Box\phi + F_2 {}^{(4)}R + F_3 {}^{(4)}G_{\mu\nu} \phi^{;\mu\nu} + \sum_{i=1}^5 A_i L_i^{(2)} + \sum_{i=1}^{10} B_i L_i^{(3)} } ,   \ee
where $F_i,A_i,B_i$ are general functions of $\phi, X$ and $ {}^{(4)}R$ is the four-dimensional
Ricci scalar.
The terms that are quadratic in second derivatives are
\begin{align} \label{ADM2}
L_1^{(2)} &= \phi_{;\mu\nu}\phi^{;\mu\nu} 
= \f{\dot\phi^2}{N^2} (\mS + \beta^2 - 2 \al ) ,\notag\\ 
L_2^{(2)} &= (\Box\phi)^2 
= \f{\dot\phi^2}{N^2} (- K + \beta)^2 , \notag\\
L_3^{(2)} &= (\Box\phi) \phi^{;\mu} \phi_{;\mu\nu} \phi^{;\nu}
= \f{\dot\phi^4}{N^4} \beta (K - \beta) , \notag\\
L_4^{(2)} &= \phi^{;\mu} \phi_{;\mu\nu} \phi^{;\nu\rho} \phi_{;\rho}
= \f{\dot\phi^4}{N^4} (\al - \beta^2), \notag\\
L_5^{(2)} &= (\phi^{;\mu} \phi_{;\mu\nu} \phi^{;\nu} )^2
= \f{\dot\phi^6}{N^6} \beta^2 , 
\end{align}
and those that are cubic are
\begingroup
\allowdisplaybreaks
\begin{align} \label{ADM3}
L_1^{(3)} &= (\Box\phi)^3 
= \f{\dot\phi^3}{N^3} (- K + \beta)^3, \notag\\
L_2^{(3)} &= (\Box\phi) \phi_{;\mu\nu}\phi^{;\mu\nu} 
= \f{\dot\phi^3}{N^3} (- K + \beta) (\mS + \beta^2 - 2 \al ), \notag\\
L_3^{(3)} &= \phi_{;\mu\nu}\phi^{;\nu\rho}{\phi^{;\mu}}_{;\rho}
= \f{\dot\phi^3}{N^3} ( - \T + 3 \V + \beta^3 - 3 \al\beta ), \notag\\
L_4^{(3)} &= (\Box\phi)^2 \phi^{;\mu} \phi_{;\mu\nu} \phi^{;\nu}
= - \f{\dot\phi^5}{N^5} \beta (- K + \beta)^2 , \notag\\
L_5^{(3)} &= (\Box\phi) \phi^{;\mu} \phi_{;\mu\nu} \phi^{;\nu\rho} \phi_{;\rho} 
= \f{\dot\phi^5}{N^5} (- K + \beta) (\al - \beta^2), \notag\\
L_6^{(3)} &= \phi_{;\mu\nu}\phi^{;\mu\nu} \phi^{;\rho} \phi_{;\rho\sigma} \phi^{;\sigma} 
= \f{\dot\phi^5}{N^5} \beta (- \mS - \beta^2 + 2 \al ) , \notag\\
L_7^{(3)} &= \phi^{;\mu} \phi_{;\mu\nu} \phi^{;\nu\rho} \phi_{;\rho\sigma} \phi^{;\sigma} 
= \f{\dot\phi^5}{N^5} ( - \V - \beta^3 + 2 \al\beta ), \notag\\ 
L_8^{(3)} &= \phi^{;\mu} \phi_{;\mu\nu} \phi^{;\nu\rho} \phi_{;\rho} \phi^{;\sigma} \phi_{;\sigma\xi} \phi^{;\xi} 
= \f{\dot\phi^7}{N^7} \beta ( \beta^2 - \al )  , \notag\\
L_9^{(3)} &= \Box\phi ( \phi^{;\mu} \phi_{;\mu\nu} \phi^{;\nu} )^2
= \f{\dot\phi^7}{N^7} \beta^2 (- K + \beta) , \notag\\
L_{10}^{(3)} &= ( \phi^{;\mu} \phi_{;\mu\nu} \phi^{;\nu} )^3
= - \f{\dot\phi^9}{N^9} \beta^3,
\end{align}
\endgroup
where we have used \eqref{fmn}  to establish the correspondence with the ADM variables.
Similarly, we can relate the coupling to the metric using the 
Gauss-Codazzi relation and integration by parts to rewrite up to boundary terms
(see e.g.\ \cite{Gourgoulhon:2007ue,Gleyzes:2013ooa})

\begin{align} \label{GCrel}
\int d^4x \sqrt{-g} F_2 {}^{(4)}R 
={} & \int d^4x \sqrt{-g} \kk{ F_2 \mk{R+K^i{}_jK^j{}_i-K^2} 
  -2\mk{ F_{2\phi} \f{  \dot\phi}{N}K + {2} F_{2X} \f{\dot\phi^2}{N^2}(\beta K-\alpha) }  } ,\\
\int d^4x \sqrt{-g} F_3 {}^{(4)}G_{\mu\nu} \phi^{;\mu\nu}
={} & \int d^4x \sqrt{-g} \Bigg[ \F_3 \f{\dot\phi}{N} \mk{ \f{KR}{2} - {K^i}_j{R^j}_i  }  + \frac{\F_{3\phi}-F_{3\phi}}{2} \f{\dot\phi^2}{N^2}R +   \frac{F_{3\phi}}{2}  \f{\dot\phi^2}{N^2}  ({K^i}_j{K^j}_i-K^2)   \notag\\
&+
F_{3X} \f{\dot\phi^3}{N^3} \mk{ ({K^i}_j{K^j}_i-K^2 )\beta  + 2 \alpha K
- 2 \V) } \Bigg],
\end{align}    
where
\be F_{3X} = \F_{3X} + \f{\F_3}{2X} . \ee

While in general, the appearance of $\alpha^i_{\hphantom{i}j}$ and $\beta$ in the Lagrangian signals
an extra degree of freedom since the lapse and shift no longer obey constraint equations,
this general Lagrangian (\ref{fullDHOST}) contains classes that propagate only 3 degrees of freedom and avoids  Ostrogradsky ghosts.
First, there is the GLPV class which defines a special relationship between the
$A_i,B_i,F_i$ coefficients 
\begin{align} \label{GLPV} 
A_1 &= -A_2 = 2F_{2X}+XF_4, \quad A_3=-A_4=2F_4, \notag\\ 
B_1 &= -\frac{B_2}{3}  = \frac{B_3 }{2} = \frac{ F_{3X}}{3}  + X F_5, \quad
-2B_4=B_5=2B_6=-B_7=6F_5, \notag\\
A_5 &= B_8=B_9=B_{10}=0 .
\end{align}
Note that the $F_4$ and $F_5$ terms are also arbitrary functions of $\phi,X$.   In the
Horndeski subclass of GLPV, where the scalar field equations themselves are explicitly second order, $F_4=F_5=0$
and the remaining functions are more typically labeled $(G_2,G_3,G_4,G_5) =(F_0,F_1,F_2,F_3)$.
It is easy to verify through Eqs.~(\ref{ADM2},\ref{ADM3},\ref{GCrel}) that this relationship eliminates the dependence of $\alpha^i_{\hphantom{i}j}$ and $\beta$ in the GLPV and Horndeski Lagrangians
leaving the EFT Lagrangian of the form $L (N, {K^i}_j, {R^i}_j;t)$.
More generally, $\alpha^i_{\hphantom{i}j}$ and $\beta$ can appear in a Lagrangian which still only propagates
3 degrees of freedom  if the functions $A_i,B_i,F_i$ satisfy a certain set of degeneracy conditions~(see \eqref{DC1-DHOST}, \eqref{DC2-DHOST} and \cite{Langlois:2015cwa,BenAchour:2016fzp}). This is the DHOST class of models.
The Lagrangian \eqref{LagEFT} also includes the scordatura degenerate theory~\cite{Motohashi:2019ymr} with a weak violation of the degeneracy condition.

In \S \ref{sec:scalarperts} we generalize these degeneracy conditions to the full 
EFT Lagrangian \eqref{LagEFT}.   Since there the dependence on ${\alpha^i}_j,\beta$ is arbitrary it encompasses theories with further higher-order products of $\phi_{;\mu\nu}$ beyond \eqref{fullDHOST}.
The Lagrangian \eqref{LagEFT} thus can represent any 
fully covariant or Lorentz-violating
theory involving up to second derivatives of metric and scalar field in the unitary gauge.  
Furthermore in comparison to EFTs that are explicitly built to encompass DHOST,
it allows terms like $\beta R$ that would only appear with different couplings between
the field and the metric than represented in \eqref{fullDHOST} (cf.~\cite{Langlois:2015cwa,BenAchour:2016fzp,Langlois:2017mxy}).

\subsection{Background and quadratic Lagrangian}%%%%%%%%%%%%%%%%%%%%%
\label{sec:lag-2}

We next consider the expansion of the ADM EFT Lagrangian \eqref{LagEFT} to quadratic
order in metric perturbations  around a spatially flat Friedmann-Lema\^itre-Robertson-Walker (FLRW) background
\be \bar N=1, \quad\bar N^i=0 ,\quad\bar h_{ij} = a^2 \delta_{ij},  \ee
for which 
\be {\bar K^i}_{\hphantom{i}j}  = H{\delta^i}_j , \quad {{ \bar R}^i}_{\hphantom{i}j} = 0, \quad {\bar \alpha^i}_{\hphantom{i} j} = 0 , \quad \bar \beta = -\f{\ddot\phi}{\dot\phi}  , \ee
where $H\equiv d\ln a/dt$ is the Hubble parameter.
Following \cite{Motohashi:2017gqb} we define the Taylor coefficients as
\begin{align} \label{cxynot} 
L \Big|_{\rm b} &= \C, \notag\\
\f{\pa L}{\pa {Y^i}_j} \Big|_{\rm b} &= \C_Y {\delta^j}_i, \notag\\
\f{\pa^2 L}{\pa {Y^i}_j \pa {Z^k}_\ell} \Big|_{\rm b} &= \C_{YZ} {\delta^j}_i {\delta^\ell}_k + \f{\tilde\C_{YZ}}{2} ( {\delta^\ell}_i {\delta^j}_k + \delta_{ik} \delta^{j\ell} ) ,  
\end{align}
where ``b'' denotes that the quantities are evaluated on the background,  $Y,Z \in \{ N, K, R, \alpha, \beta \}$ and the index structure is determined by the symmetry of the background.  
For notational simplicity we treat scalars and traces with the same notation; thus implicitly $ {N^i}_i \equiv N$, ${\beta^i}_i \equiv \beta$ and $\tilde\C_{NZ} = \tilde\C_{\beta Z} = 0$.

We can further eliminate terms linear in $\delta K=K-3H$ through the identity
\be \int d^4 x \sqrt{-g} F(t) K 
= -\int d^4 x \sqrt{-g} n^\mu F_{;\mu} 
= - \int d^4 x \sqrt{-g} \f{\dot F}{N} , \ee
which follows from $K={n^\mu}_{;\mu}$ ignoring boundary terms.  
The Lagrangian ${\cal L} = N\sqrt{h}L$ up to quadratic order in metric perturbations becomes
\be \label{LagEFT3} {\cal L} = N\sqrt{h} ( \C- 3 H \C_K) - \sqrt{h} \dot\C_K 
+ N\sqrt{h} (\C_N\delta N + \C_R\delta R + \C_\alpha\delta\alpha + \C_\beta\delta\beta) 
+ \f{a^3}{2}  \sum_{Y, Z} (\C_{YZ} \delta Y\delta Z + \tilde\C_{YZ} \delta {Y^i}_j \delta {Z^j}_i). \ee
In this form  $\delta K$ only shows up in the quadratic-order terms, and we need only its first-order perturbation.
In contrast, we expand ${\delta R^i}_j, \delta {\alpha^i}_j, \delta \beta$ up to quadratic order
\be
\delta {Y^i}_j = \delta_1 {Y^i}_j + \delta_2 {Y^i}_j + \cdots .
\ee
We note that $\delta_1 {R^i}_j,\delta_2 {R^i}_j$ do not involve $\delta N,\delta N^i$,
whereas
\begin{align} \label{dadb}
\delta_1 {\alpha^i}_j &= 0, \notag\\
\delta_2 {\alpha^i}_j &= \f{1}{a^2} \delta^{ik} \delta N_{,j}\delta N_{,k},\notag\\
\delta_1 \beta &= -\bar\beta \delta N +  \dot{\delta N}, \notag\\
\delta_2 \beta &= \bar\beta \delta N^2  - 2 \delta N \dot{\delta N} - \delta N^i \delta N_{,i} ,
\end{align}
since
\begin{align}
n^\mu &\approx (1- \delta N + (\delta N)^2, - \delta N^i (1-\delta N)), \notag\\
(\ln N)_{,\mu} &\approx \delta N_{,\mu} (1 - \delta N) ,
\end{align}
up to quadratic order.
Note that for the perturbed FLRW metric, ${\alpha^i}_j$ is a quadratic-order quantity.
The background equations are given by varying the action with respect to $N$ and $\sqrt{h}$
\begin{align} \label{bgeq}
\C - 3 H\C_K + \C_N - (3H+\bar\beta) \C_\beta - \dot \C_\beta &= 0, \notag\\
\C - 3H\C_K - \dot \C_K &= 0 .
\end{align}

With the background equation and integration by parts, the quadratic Lagrangian becomes
\begin{align} \label{L2}
{\cal L}_2 
&= a^3 \left[ 
\C_R \mk{ \delta_1 R \f{\delta_1\sqrt{h}}{a^3} + \delta_2 R }
- \C_\beta \delta N \mk{ \f{\delta_1\sqrt{h}}{a^3} }^\centerdot   
- \C_\beta \delta N^i \delta N_{,i}
+ \f{1}{2} \C_{\beta\beta} \dot{\delta N}^2 
+ \C_\alpha \f{1}{a^2}\delta^{ij} \delta N_{,i} \delta N_{,j}  \right.
\notag\\
&~~~
+ \f{1}{2} \mk{2 \C_N + \C_{NN} { - 2 \bar\beta \C_{\beta N} + \bar\beta^2\C_{\beta\beta} } + \f{\mk{a^3\C_\beta}^\centerdot}{a^3} - \f{\mk{a^3\C_{\beta N}}^\centerdot}{a^3} { + \f{\mk{a^3\bar\beta\C_{\beta\beta}}^\centerdot}{a^3} }  } \delta N^2
+ ( \C_{\beta K} \delta_1 K + \C_{\beta R} \delta_1 R ) \dot{\delta N} \notag\\
&~~~ + \left. { [ (\C_{NK} -\bar\beta\C_{\beta K}) \delta_1 K + (\C_{NR} + \C_R-\bar\beta\C_{\beta R}) \delta_1 R ] \delta N }
+ \f{1}{2}\sum_{Y=K,R}\sum_{Z=K,R} (\C_{YZ}\delta_1 Y\delta_1 Z + \tilde\C_{YZ} \delta_1 Y^i_j \delta_1 Z^j_i )  \right] .
\end{align}
This expansion can be continued to higher order for the computation of non-Gaussianity 
(see e.g.~\cite{Passaglia:2018afq}).

\subsection{Scalar perturbations}%%%%%%%%%%%%%%%%%%%%%
\label{sec:scalarperts}

From the quadratic Lagrangian~\eqref{L2}, we note that the new terms $\C_{\alpha}, \C_{\beta}, \C_{\beta Y}$ are always accompanied by $\delta N$, which is a natural consequence of \eqref{dadb}.
Therefore, ${\alpha^i}_j$ and $\beta$ dependencies of the Lagrangian change the dynamics of scalar
but not vector or tensor perturbations which are given explicitly in \cite{Motohashi:2017gqb}.

For scalar perturbations
\be \label{spert} N = 1+\delta N, \quad N_i = \pa_i \psi, \quad h_{ij} = a^2e^{2\zeta} \delta_{ij} .\ee
Following \cite{Motohashi:2017gqb}, we use 
\begin{align}
\delta \sqrt{h} &= 3a^3 \zeta, \notag\\
\delta  K^i_{\hphantom{i}j} &= (\dot\zeta-H\delta N)\delta^i_{\hphantom{i}j} -\f{1}{a^2}\delta^{ik}\pa_k\pa_j \psi, \notag\\
\delta  K &= 3(\dot\zeta-H\delta N) - \f{\pa^2 \psi}{a^2}, \notag\\
\delta_1 R^i_{\hphantom{i}j} &= - \f{1}{a^2} ( \delta^i_{\hphantom{i}j} \pa^2 \zeta + \delta^{ik} \pa_k \pa_j \zeta) ,\notag\\ 
\delta_2 R &= -\f{2}{a^2} [(\pa\zeta)^2-4\zeta \pa^2\zeta ] 
\sim -\f{10}{a^2} (\pa\zeta)^2,
\end{align}
where the last equality for $\delta_2 R$ holds up to a total derivative,
to obtain the quadratic Lagrangian 
in Fourier space as 
\begin{align} \label{L2s-c}
\L_2 &= \f{1}{2}c_1 \dot\zeta^2 + c_2 \dot\zeta \dot{\delta N} + \f{1}{2}c_3 \dot{\delta N}^2 + \mk{c_4 + c_5 \f{k^2}{a^2}} \dot\zeta \delta N \notag\\
&~~~ + \f{1}{2}\mk{c_6 + c_7 \f{k^2}{a^2}} \f{k^2}{a^2} \zeta^2 + c_8 \f{k^2}{a^2} \zeta \delta N + \f{1}{2}\mk{c_9 + c_{10} \f{k^2}{a^2}} \delta N^2 \notag\\
&~~~ + \f{1}{2}c_{11} \f{k^4}{a^4} \psi^2 + \f{k^2}{a^2} \psi \mk{ \f{c_1}{3} \dot\zeta + \f{c_2}{3} \dot{\delta N} + \f{c_4}{3} \delta N + c_{12} \f{k^2}{a^2} \zeta } ,
\end{align}
where
\begin{align} \label{coeffs-2}
c_1 &= 3a^3 (3\C_{KK}+\tilde\C_{KK}), &
c_2 &= 3 a^3 \C_{\beta K}  , &
c_3 &= a^3\C_{\beta\beta} , &
c_4 &= -3a^3 \Theta,&\notag\\
c_5 &= - 4a^3\C_{\beta R} ,&
c_6 &= 4 a^3 \Psi ,&
c_7 &= 2a^3(8\C_{RR} + 3\tilde\C_{RR}),&
c_8 &= 4 a^3 \Xi, &\notag\\
c_9 &= a^3 \Phi  ,&
c_{10} &= 2 a^3 \C_\alpha ,&
c_{11} &= a^3 (\C_{KK}+\tilde\C_{KK})  , &
c_{12} &= 2a^3 (2\C_{KR}+\tilde\C_{KR}) ,
\end{align}
and
\begin{align} 
\Phi &\equiv 
   2 \C_N + \C_{NN}  { - 2 \bar\beta \C_{\beta N} + \bar\beta^2\C_{\beta\beta} } + ( \C_\beta - \C_{\beta N}  +\bar\beta\C_{\beta\beta})^\centerdot \notag\\
   &~~~+ 3 H [\C_{\beta} - \C_{\beta N}  { +\bar\beta\C_{\beta\beta} } - 2  {(\C_{NK} -\bar\beta\C_{\beta K})} + \dot\C_{\beta K} ] 
   + 3 \C_{\beta K} \dot H 
   + 3 H^2 [3 (\C_{\beta K} + \C_{KK}) + \tilde\C_{KK}],\notag\\
\Psi &\equiv 
   \C_R - 3 \dot\C_{KR} - \dot{\tilde\C}_{KR} - 
   H (3 \C_{KR} + \tilde\C_{KR}),\notag\\
\Xi &\equiv 
    {\C_{NR} + \C_R-\bar\beta\C_{\beta R}} - \dot\C_{\beta R} - 
   H (\C_{\beta R} + 3 \C_{KR} + \tilde\C_{KR}),\notag\\
\Theta &\equiv \C_{\beta} -  {(\C_{NK} -\bar\beta\C_{\beta K})} + H (3 \C_{KK} + \tilde\C_{KK}).
\end{align}
We highlight these four combinations as they involve time derivatives or the Hubble parameter and degeneracy conditions involving them would typically need to arise from integration by parts on the Lagrangian [see e.g.~(\ref{dKdRint})].

In Appendix \ref{app:deg}, generalizing Ref.~\cite{Langlois:2017mxy}, we provide a complete analysis of the construction of degeneracy conditions imposed on the various $c_i$ coefficients which we briefly summarize here.
The result is 8 types of degeneracy conditions, cases $1a \ldots 3 c$ (with 3a impossible to satisfy), each of which may be realized by the $c_i$ or equivalently the $\C_i$ coefficients in various ways.   
The degeneracy conditions we derive apply 
for any theory involving second-order derivatives in any form 
in the Lagrangian for unitary gauge~\eqref{LagEFT}.
Since the Lagrangian~\eqref{LagEFT} allows any dependence on 
$(N,{K^i}_j,{R^i}_j,{\alpha^i}_j,\beta)$, 
or equivalently on second derivatives \eqref{fmn}, 
this degeneracy conditions applies beyond 
the quadratic and cubic DHOST theories.
Furthermore, in general it also applies to Lorentz-violating theories.

The first condition required for the single scalar propagating degree of freedom is degeneracy in the temporal structure of (\ref{L2s-c}).  
Of the three possibilities, we focus on the case 1 type where the condition $c_3 = c_2/c_1^2$
is satisfied and the combination 
\begin{equation}
\tilde \zeta = \zeta + \frac{c_2}{c_1}\delta N,
\end{equation}
alone carries the temporal derivatives.  
 The other two cases have the lapse $\delta N$ as the propagating degree of freedom and would cause difficulties in
recovering an observationally viable theory of gravity.    More generally our linear degeneracy conditions should be viewed
as necessary, but not necessarily sufficient, conditions for a viable nonlinear scalar-tensor theory of gravity.

Under this  $c_3=c_2/c_1^2$ condition, the quadratic Lagrangian~\eqref{L2s-c} for scalar perturbation in unitary gauge would appear to propagate only 1 degree of freedom.
This degeneracy condition  applies to the scalar quadratic Lagrangian 
in any theory involving second-order derivative of any form 
in Lagrangian in the unitary gauge~\eqref{LagEFT} and includes the DHOST models as well as the Horndeski or GLPV models where $c_2=0$ and $\tilde\zeta=\zeta$.

However, the temporal degeneracy condition alone is not sufficient to guarantee that there is only a single degree of freedom.   
In terms of the Euler-Lagrange equations, it only removes the fourth-order derivatives and third-order derivatives still need to be removed to avoid unbounded Hamiltonian~\cite{Motohashi:2014opa}.
Furthermore, 
if unitary gauge defines a foliation that corresponds to characteristic surfaces of the second degree of freedom then its dynamics 
are hidden from this temporal structure.
Since such a degree of freedom propagates instantaneously on this  surface, 
it is not a Cauchy surface upon which initial conditions can be propagated forwards in time.   Hence its temporal kinetic terms vanish.   However on a noncharacteristic surface, temporal kinetic terms reappear and can possess a well-posed Cauchy problem 
as discussed in detail in~\cite{Motloch:2015gta,Motloch:2016msa}.   One should therefore not take the apparent lack of an extra degree of freedom in unitary gauge as a definitive absence 
(cf.\ \cite{Gao:2014soa}). Of course, the counting of degrees of freedom
cannot depend on the gauge or ADM slicing and so we expect additional degeneracy conditions that involve the spatial derivatives of the kinetic matrix in unitary gauge.

For a $1+1$ dimensional system of linear partial differential equations, including the plane parallel Fourier modes considered below,
one can exploit the algorithm~\cite{Motloch:2016msa} based on the Kronecker form of a matrix pencil which  includes all possible linear combinations of temporal and spatial derivatives to count degrees of freedom and find characteristic curves in the presence of any hidden constraints
(see Appendix of \cite{Motloch:2016msa}).
However for the quadratic and cubic DHOST theories, the full covariant and nonlinear degeneracy conditions are already known.  As shown in the Appendix, we can obtain the remaining conditions for the
quadratic action by demanding that the dispersion relation of remaining degree of freedom
take  their normal linear form in unitary gauge.  This logic  also applies to the wider class of degenerate theories that originate from a covariant action and so we retain terms that are absent in the quadratic and cubic DHOST Lagrangian in Appendix~\ref{app:deg}.

We now focus in particular on the degeneracy conditions given in (\ref{D1}) in case 1a, as other branches may not have phenomenologically viable theories of gravity associated with them.   We emphasize though that this same procedure
can be carried out for any of the branches.    In this case, the conditions on the $c_i$ coefficients are
\begin{equation}
\label{ccond1a}
c_3 = \f{c_2^2}{c_1},\quad c_5=c_7=c_{11}=c_{12}=0, \quad
c_{10} = 2c_8 x - c_6 x^2 , 
\end{equation}
where $x=c_2/c_1$, and these conditions imply
\begin{align}
\label{eqn:degcon2NIIa}
\C_{\beta\beta} = \f{3 \C_{\beta K}^2}{3\C_{KK}+\tilde\C_{KK}}  , \quad
\tilde \C_{KK} &= -\C_{KK}, \quad \tilde\C_{KR}=-2 \C_{KR} , \quad \tilde \C_{RR}=-\frac{8}{3} \C_{RR} ,\quad
\C_\alpha  = 
\frac{2 \C_{\beta K}}{ \C_{KK}}  \Xi
- \frac{ \C_{\beta K}^2}{2  \C_{KK}^2} \Psi , \quad \C_{\beta R} =0  .
\end{align}
This branch includes the $^2$N-I/Ia class of quadratic and cubic DHOST, GLPV and Horndeski theories.

Under these conditions we can simplify \eqref{L2s-c} as
\begin{align} \label{L2s-d}
\L_2 &= \f{1}{2}c_1 \dot{\tilde\zeta}^2 + \mk{c_4-c_1\dot x } \dot{\tilde\zeta} \delta N  
+ \f{1}{2}c_6 \f{k^2}{a^2} \tilde\zeta^2 
+ (c_8-c_6x) \f{k^2}{a^2} \tilde\zeta \delta N 
\notag\\ &~~~
+ \f{1}{2}\mk{  c_9 + \dot c_4 x + c_1 \dot x^2 - c_4 \dot x} \delta N^2 
+ \f{k^2}{a^2} \psi \mk{ \f{c_1}{3} \dot{\tilde\zeta} + \f{c_4-c_1\dot x}{3} \delta N  } .
\end{align}
The equation of motion for $\psi$ and $\delta N$ yields the constraints
\begin{align} 
\delta N &= \f{c_1}{c_1\dot x- c_4} \dot{\tilde\zeta} ,\notag\\
\f{k^2}{a^2}\psi &= \f{3}{c_1\dot x - c_4}\kk{ \mk{c_4-c_1\dot x } \dot{\tilde\zeta}
+ (c_8-c_6x) \f{k^2}{a^2} \tilde\zeta
+ \mk{c_9 + \dot c_4 x + c_1 \dot x^2 - c_4 \dot x} \delta N },
\label{constraint}
\end{align}
where we have assumed 
\be \label{c1dxmc4} 0 < |\Omega|< \infty, \ee
with $\Omega \equiv c_1\dot x- c_4$, which generalizes 
the condition $2H\C_{KK}\ne \C_{NK}$ employed in Eq.~(33) of \cite{Motohashi:2017gqb} to
cases where the Lagrangian depends $\alpha^i{}_j, \beta$.
Violation of this condition makes unitary gauge perturbations ill-defined. 
For singular $\Omega$, the kinetic term vanishes and hence the system is strongly coupled.
On the other hand, for $\Omega= 0$, unitary gauge itself is ill-defined.
To see this, we follow \cite{Motohashi:2017gqb} and move
to a comoving gauge defined by the condition that for the perturbed Einstein tensor $\delta G^0{}_i=0$  for a general metric theory of gravity  \cite{Hu:2016wfa}. 
The gauge transformation from  unitary gauge to comoving gauge is characterized by the time shift
 $T = -\Delta/\dot H$, where $\Delta \equiv H\delta N - \dot \zeta$ (see Eq.~(B14) of \cite{Motohashi:2017gqb}).
Using \eqref{constraint}, we have
\be \Delta = \f{1}{\Omega} \ck{ \kk{c_1(Hx-\dot x)+c_4}\dot \zeta 
+ Hc_1x (\dot{\tilde\zeta} - \dot \zeta) }, \ee
so that $\Omega=0$ makes $\Delta$ diverge, implying that the gauge transformation between the two gauges requires an infinite time shift and hence is ill-defined.  
Note that this is not necessarily a problem if the original system of
equations in $(\delta N, \tilde\zeta ,\psi)$ possesses only regular singular points
and  is hence integrable without first imposing the constraint equation (see \cite{Ijjas:2017pei,Dobre:2017pnt} for a related discussion).  
Furthermore if $\tilde\zeta$ freezes out but $\tilde\zeta-\zeta$ continues
to evolve outside the horizon, then the two gauges will differ.   
We construct an explicit model where this occurs in \S\ref{sec:inf}
(see \cite{Lagos:2019rfc} for a discussion of related cases).

Note also that if $c_8 = c_6x$ the $k^2 \tilde \zeta$ term vanishes in the Euler-Lagrange equation \eqref{constraint} for $\delta N$, and prevents the recovery of Newtonian gravity for nonrelativistic matter, as found in \cite{Langlois:2017mxy} for one of the DHOST subclasses (see Appendix~\ref{app:deg} for more details).

Substituting the constraints (\ref{constraint}) into the Lagrangian~\eqref{L2s-d} and integrating by parts
give the usual Mukhanov-Sasaki form for the quadratic Lagrangian 
\be \label{S2s} {\cal L}_2 = A_\zeta \dot{\tilde\zeta}^2 - B_\zeta \frac{k^2}{a^2} \tilde\zeta^2 , \ee
where
\begin{align}
A_\zeta ={}&\frac{c_1}{2}  \frac{  c_1(c_9 +  \dot c_4 x) + c_4 ( c_1 \dot x- c_4 ) }{  ( c_1 \dot x- c_4)^2},
\nonumber\\
B_\zeta={}& a^2  \left( \frac{c_1}{2 a^2} \frac{ c_8 - c_6 x}{ c_1\dot x- c_4} \right)^\centerdot - \frac{c_6}{2}.
\end{align}
From these terms, we can define the scalar sound speed $c_s^2$ and the normalization
parameter $b_s$ as
\be \label{bscsA}
c_s^2 = \f{B_\zeta}{A_\zeta}, \quad 
b_s = \f{B_\zeta}{a^{3}\epsilon_H}. \ee
For a canonical scalar field $c_s^2=b_s=1$.
These expressions are generalizations of Eq.~(37) of \cite{Motohashi:2017gqb}.

\section{D-Inflation with time varying EFT coefficients}%%%%%%%%%%%%%%%%%%%%%%%%%%%%%%%%%%%%%%%%%
\label{sec:inf}

In this section we consider models of degenerate higher-order inflation (D-inflation)  with time varying EFT coefficients, specifically in the quadratic DHOST class.     In general, EFT coefficients can vary in time 
and  one needs to evaluate carefully the slow-roll hierarchy of all dynamical parameters, for which the generalized slow-roll approximation developed in \cite{Motohashi:2017gqb}  provides a systematic framework, based on the
evolution of $H$, $\e_H$, $b_s$ and  $c_s$ in \eqref{bscsA} as well as the analogous quantities for tensors
\be \label{btct} b_t = 2\C_R , \quad c_t^2 = \f{2\C_R}{\tilde\C_{KK}} . \ee
D-inflation provides an additional motivation for these time-varying considerations
in that one might seek to construct models where their novel features are only present
during inflation and are absent thereafter where they would otherwise impact 
cosmological and astrophysical observables.
We construct our model in \S\ref{ssec:3A}, and elucidate the novel features on background dynamics and evolution of perturbations in \S\ref{ssec:3B} and \S\ref{ssec:3C}, respectively.

\subsection{D-inflation model}%%%%%%%%%%%%%%%%%%%%%
\label{ssec:3A} 

As a concrete example,
let us require the inflationary model to satisfy $c_t^2-1=0$ and $\tilde \zeta - \zeta=0$ at the end of inflation. 
The former is the requirement for tensor sound speed  to be the speed of light, imposed at least at the present epoch by observation of gravitational waves from binary neutron star merger and its electromagnetic counterpart~\cite{TheLIGOScientific:2017qsa,Monitor:2017mdv}.
The latter is imposed as we would like inflation to become fully canonical so that by reheating everything is as usual.
We shall see that enforcing this requirement for all perturbation quantities allows us to avoid instabilities caused by derivative couplings~\cite{Ramirez:2018dxe}.

We can concretely implement these requirements using the EFT of a quadratic DHOST 
model starting with the case 1a degeneracy conditions \eqref{eqn:degcon2NIIa}.
Since $\tilde\C_{KR}=\C_{KR}=\tilde\C_{RR}=\C_{RR}=0$ in this case, the third and the fourth conditions in \eqref{eqn:degcon2NIIa} identically hold.
From the second condition in \eqref{eqn:degcon2NIIa} we obtain
\be \label{DC1-DHOST} A_1 = -A_2 .\ee
Plugging it into the first and fifth conditions in 
\eqref{eqn:degcon2NIIa} and solving the two equations for $A_4$ and $A_5$, we obtain
\begin{align} \label{DC2-DHOST}
A_4 &= \f{2(A_2+2 F_{2X})}{X} - \f{(2 A_2 + 4 F_{2X} + X A_3 ) [8 F_2^2 + 2 X A_2 (5 F_2 - 8 X F_{2X} ) + X^2 A_3 F_2 - 12 X F_2 F_{2X} ]}{8 X (F_2 + X A_2 )^2}  ,\notag\\
A_5 &= \f{(2 A_2 + 4 F_{2X} + X A_3 ) [4 A_3 F_2 - A_2 (2 A_2 + 4 F_{2X} - 3 X A_3 )]}{8 (F_2 + X A_2 )^2} ,
\end{align}
which matches Eqs.~(5.1) and (5.2) in \cite{Langlois:2015cwa}. 
In general a degenerate theory in this class is identified by 
$A_2,A_3,F_2$ and  one can choose $F_0$, $F_1$ as free functions without affecting the degeneracy structure.
Note that if $\C_{\beta K}=0$ then 
\be 2 A_2 + 4 F_{2X} + X A_3 = 0 ,\ee 
so that 
$A_2$ and $A_3$ are no longer independent; 
this case corresponds to Eq.~(5.3) in \cite{Langlois:2015cwa} and reproduces the GLPV restriction for quadratic terms in \eqref{GLPV}.

Next, from \eqref{ADM2}, we obtain the tensor sound speed as 
\be \label{ctcon} c_t^2 - 1 = \f{2\C_R}{\tilde \C_{KK}} - 1 = \f{XA_1}{F_2-XA_1} = -\f{XA_2}{F_2+XA_2}, \ee
where we used \eqref{DC1-DHOST} and
\be \label{zetacon} \tilde\zeta-\zeta = \f{\C_{\beta K}}{3\C_{KK}+\tilde\C_{KK}} 
= - \f{X (2 A_2 + 4 F_{2X} + X A_3 )}{4 F_2 + 2 X (A_1 + 3 A_2) }.
\ee
For the Horndeski theory, requiring the right-hand side of \eqref{ctcon} to vanish implies that $F_2=F_2(\phi)$.  
Note also that the right-hand side of \eqref{zetacon} identically vanishes for Horndeski and GLPV theories.

We would like to choose the functions $A_2,A_3,F_2$ to make these two quantities \eqref{ctcon}, \eqref{zetacon} be nonzero during inflation and evolve to zero by the end of inflation. 
As a simple example, we set 
\be F_0=-\f{X}{2}-V(\phi), \quad F_1=0, \quad F_2=\f{1}{2}, \quad A_3=0 , \ee
where we work in natural units $\Mpl\equiv (8\pi G)^{-1/2}=1$, and
for which the degeneracy conditions \eqref{DC1-DHOST}, \eqref{DC2-DHOST} yield
\be A_1=-A_2,\quad 
A_4 = \f{A_2^2 (3 + 8 X A_2)}{(1 + 2 X A_2)^2}, \quad 
A_5 = -\f{2 A_2^3}{(1 + 2 X A_2)^2} , \ee
and \eqref{ctcon}, \eqref{zetacon} read
\be c_t^2 - 1 = 2(\tilde\zeta-\zeta) = \f{1}{1+2XA_2} -1 \equiv \theta(\phi,X). \ee
Here, we are interested in a function $\theta$ such that it starts from finite value and evolves to zero 
either from the evolution of $X$ or an appropriate form for $A_2(\phi,X)$.

Under these assumptions, the action is given by 
\begin{align} \label{infmodel}
S 
&= \int d^4x N\sqrt{h} \Bigg[ \f{1}{2}(R+K^i_jK^j_i-K^2) + \f{\dot\phi^2}{2N^2} - V(\phi)
+A_2\f{\dot\phi^2}{N^2} ( K^2 - \mS - 2\beta K + 2\alpha)
 \notag\\
&~~~+\f{A_2^2}{{ 1 - 2 \f{\dot\phi^2}{N^2} A_2}} \frac{\dot\phi^4}{N^4} \left(  \frac{3 - 8 \f{\dot\phi^2}{N^2} A_2}{1 - 2 \f{\dot\phi^2}{N^2} A_2} \al - 3\beta^2 \right) \Bigg] .
\end{align}
For simplicity, we will illustrate this model with $A_2={\rm const}$, or at least nearly so during
most of the $\sim 60$ efolds before the end of inflation.   We shall see that in models where the field oscillates at reheating, $A_2$ needs to vanish
before this point to avoid gradient or ghost instabilities.
However, any late-time change does not affect large-scale observables which are well outside
the horizon at that point.

\subsection{Background dynamics}%%%%%%%%%%%%%%%%%%%%%
\label{ssec:3B}

From \eqref{infmodel} we can calculate EFT coefficients.  For instance, those which are necessary for the background equations~\eqref{bgeq} are
\begin{align} \label{C4}
\C &= \f{\dot\phi^2}{2} - V - 3 H_b^2(1-2 \dot\phi^2 A_2)   ,\notag\\
\C_K &= - 2 H_b (1- 2 \dot\phi^2 A_2) , \notag\\
\C_N &= -\dot\phi^2 - 12 H_b [H - H_b(1- \dot\phi^2 A_2)] ,\notag\\
\C_\beta &= - 6 A_2 \dot\phi^2 H_b ,
\end{align}
and those for the perturbations are
\begin{align} \label{C5}
\Xi &= \Psi = \C_R = \f{1}{2}, \qquad   
\tilde \C_{KK} = - \C_{KK} = 1 - 2\dot\phi^2 A_2 ,\qquad   
\C_{\beta K} = -2 \dot\phi^2 A_2 , \notag\\
\C_{\beta N} &= \f{12 \dot\phi^2 A_2  [2 (1 - \dot\phi^2 A_2 ) H_b - H]}{1 - 2 \dot\phi^2 A_2 }, \qquad
\C_{NK} = 4 [(1 - 2 \dot\phi^2 A_2 ) H_b - (1 + \dot\phi^2  A_2) H ] , \notag\\
\C_{NN} &= 3(1+ 12 A_2 H_b^2)\dot\phi^2 + 36 (H-H_b)H_b - \frac{24 (H-H_b)^2}{1-2 \dot\phi^2 A_2},
\end{align}
where \cite{Crisostomi:2018bsp}
\begin{equation} \label{defHb}
H_b \equiv  H - \f{A_2 \ddot\phi \dot\phi}{1 - 2 A_2 \dot\phi^2} .
\end{equation}

The background equations~\eqref{bgeq} are then given by
\begin{align}
  \label{Eq1} 6 \dot\phi^2 A_2 (\dot H_b - H H_b) + 3 (1 + 2 \dot\phi^2 A_2 ) H_b^2 - \f{1}{2} \dot\phi^2 - V = 0, \\
  \label{Eq2} 2 (1 - 2 \dot\phi^2 A_2 ) (\dot H_b - H H_b) + 5 H_b^2 - 10 \dot\phi^2 A_2 H_b^2 + \f{1}{2} \dot\phi^2 - V = 0 .
\end{align}
Note that $A_2=0$ recovers the Einstein equations in canonical inflation.
While the system involves $\dddot\phi$ via $\dot H_b$, by virtue of the degeneracy conditions, it is equivalent to a system
whose evolution is determined by
initial data in a single degree of freedom e.g.\ for the background, the position of the field in phase space $(\phi,\dot\phi)$.

The first step in establishing this equivalence is to eliminate
$\dot H_b$ from \eqref{Eq1} and \eqref{Eq2} to obtain:
\be \label{Eq3} 6 (1 - 5 \dot\phi^2 A_2 + 6 \dot\phi^4 A_2^2 ) H_b^2 - \dot\phi^2 (1 + \dot\phi^2 A_2) - 2 (1 - 5 \dot\phi^2 A_2) V = 0. \ee
Hence, there are two branches for $H_b$.  In general, if $A_2\ne{\rm const}$ we would have a term linear in $H_b$ but here, we obtain simple positive and negative roots
\be \label{Hbsols} H_b = \sigma \sqrt{\f{ \dot\phi^2 (1 + \dot\phi^2 A_2) + 2 (1 - 5 \dot\phi^2 A_2) V}{6 (1 - 5 \dot\phi^2 A_2 + 6 \dot\phi^4 A_2^2 )} } , \ee
where $\sigma = \pm 1$. 
Next, we choose one of the two branches of $H_b=H_b(\phi,\dot\phi)$ and take
its time derivative  $\dot H_b=\dot H_b(\phi,\dot\phi,\ddot\phi)$. 
Substituting $H_b=H_b(\phi,\dot\phi)$ and $\dot H_b=\dot H_b(\phi,\dot\phi,\ddot\phi)$ into \eqref{defHb} and either of \eqref{Eq1} or \eqref{Eq2}, we obtain two equations for $H=H(\phi,\dot\phi,\ddot\phi)$.
Finally eliminating $H$ from the two equations, we obtain an equation for $\ddot\phi=\ddot\phi(\phi,\dot\phi)$ governing the evolution of the system from a point in phase space.
From this evolution we can then define $H=H(\phi,\dot\phi)$ and other background quantities which
define the slow-roll parameters.
Note that equations depend only on $m^2A_2$ if one rescales time to $mt$.  Hence, so long as $m^2A_2$ is fixed, the relative evolution in $mt$ is the same for various values of $m$ with only the amplitudes $H\propto m$ and $\dot\phi\propto m$ changing.
Given the inflationary dynamics, we can check the condition~\eqref{c1dxmc4} for whether unitary gauge perturbations are well-defined.  
In our model, 
\be \label{Omega} \Omega= 6 a^3 [-H_b + \dot\phi^2 A_2 (2 H + 3 H_b)] , \ee
which should be a finite value.

Even at the background level, this procedure produces novel behavior in phase space.
First, not all phase space positions are allowed, even for a positive potential,
and allowed  positions can evolve into or from disallowed regions.
For definiteness consider the quadratic potential $V(\phi)=m^2\phi^2/2$.
With this potential, for both branches of $H_b$ in \eqref{Hbsols}, $H(\phi,\dot\phi)$ is singular at
\be \label{red-boundary} \phi=\pm \sqrt{\f{1 + \dot\phi^2 A_2 - 9 \dot\phi^4 A_2^2 + 3 \dot\phi^6 A_2^3 }{m^2A_2 (1 + 4 \dot\phi^2 A_2 - 15 \dot\phi^4 A_2^2 )}}, \quad \text{or} \quad \f{\dot\phi}{m}=\pm\f{1}{\sqrt{3m^2A_2}} .\ee
For instance, plugging $\f{\dot\phi}{m}=\f{1}{\sqrt{3m^2A_2}}-\delta$ with an infinitesimal variable $\delta$ into $H(\phi,\dot\phi)$ and Taylor expanding around $\delta=0$ yields  
\be \label{Hsing} \f{H}{m} = 3^{-5/4} (m^2A_2)^{-3/4}\sqrt{1 - \f{3}{2} m^2A_2 \phi^2} \f{\sigma}{\sqrt{\delta}} + \O (\delta^{1/2}) , \ee
which is indeed singular at $\delta=0$ for both $\sigma=\pm 1$ branches so long as $1 - \f{3}{2} m^2A_2 \phi^2 \ne 0$.
For $1 - \f{3}{2} m^2A_2 \phi^2 > 0$, the Hubble parameter is real for the $\delta>0$ side of the boundary, and imaginary for the $\delta<0$ side.  
On the other hand $H={\O(\delta^{-1})}\rightarrow \pm \infty$ on alternate sides of the $\phi$ values of the first case in \eqref{red-boundary}.

Also, there are boundaries across which $H(\phi,\dot\phi)$ changes from real to complex value while remaining finite:
\be \label{dashed-red-boundary} \phi=\pm \f{\dot\phi}{m} \sqrt{\f{ 1 + \dot\phi^2 A_2 }{ -1 + 5 \dot\phi^2 A_2 }},\quad \text{or} \quad \f{\dot\phi}{m}=\pm\f{1}{\sqrt{2m^2A_2}} , \ee
where $H_b$ is zero or singular respectively.
For instance, plugging $\f{\dot\phi}{m}=\f{1}{\sqrt{2m^2A_2}}+\delta$ into $H(\phi,\dot\phi)$ and $\ddot\phi(\phi,\dot\phi)$ and Taylor expanding around $\delta=0$ yields 
\begin{align}  \label{taylor2}
\f{H}{m} &= \f{\sqrt{2m^2A_2} \phi}{1 - 2 m^2A_2 \phi^2} +  \f{2^{3/4} (7 - 10 m^2A_2 \phi^2)  }{3 (m^2A_2)^{1/4}\sqrt{1 - 2 m^2A_2 \phi^2} } \sigma \sqrt{\delta} + \O(\delta) , \nonumber\\
\frac{\ddot \phi}{m^2} &= \frac{2^{1/4} }{(m^2 A_2 )^{3/4} } \sqrt{1 - 2 m^2 A_2\phi^2}\sigma \sqrt{\delta}
+\O(\delta).
\end{align}
Therefore, for $A_2>0$ and $1 - 2 m^2A_2 \phi^2>0$, approaching the boundary $\f{\dot\phi}{m}=\f{1}{\sqrt{2m^2A_2}}$ from the positive $\delta$ side, $H(\phi,\dot\phi)$ changes from real to complex value for both branches.
Furthermore, as can be seen from the slope $\ddot\phi/\dot\phi$,
the $\sigma=\pm 1$ trajectories form two halves of a parabola  whose minimum intersects 
the boundary.
Taylor expansion around another boundary in \eqref{dashed-red-boundary} also has a similar structure.

Another interesting point is that the branches of $H_b$ are not in general related by time reversal as they would be for $H$ in GR beyond the $A_2={\rm const}$ case, where there would be terms linear in $H_b$ in \eqref{Eq3}.  
This means that for a given initial position in the field phase space, evolution is not
unique without specifying the branch choice for the metric.   This feature is shared by the
class of Galileon or G-inflation models as well \cite{Deffayet:2010qz,Kobayashi:2011nu}.

We take an example parameter set
based on the following rough estimation. 
Proceeding backwards from the end of inflation on the $\sigma=+1$ branch when the field approaches the origin $(\phi,\dot\phi/m)=(0,0)$,
we see that since D-inflation effects for the background scale as $\dot\phi^2 A_2$, 
the background
field behaves close to the canonical model with  $\e_H \approx 2/\phi^2$ and
$\dot\phi/m \approx \sqrt{2/3}$ on the slow-roll attractor. 
Thus for the canonical phase to last at least $\sim 60$ efolds, we require $\phi\approx 15$ to be in the canonical phase.   This phase is bounded at some maximum $|\phi|$ by encountering the first of the boundaries~\eqref{red-boundary} where $H$ is singular.   For small $\dot\phi/m$ this
occurs at $\phi\approx \pm (m^2A_2)^{-1/2}$, and hence we 
require $|m^2A_2 \phi^2 | < 1$ or $m^2A_2 < 1/15^2$.
Thus, as an example, we set 
$m^2A_2=0.002$.

% ==================== Figure 1 ====================
\begin{figure}[t]
\centering
\includegraphics[width=.45\textwidth]{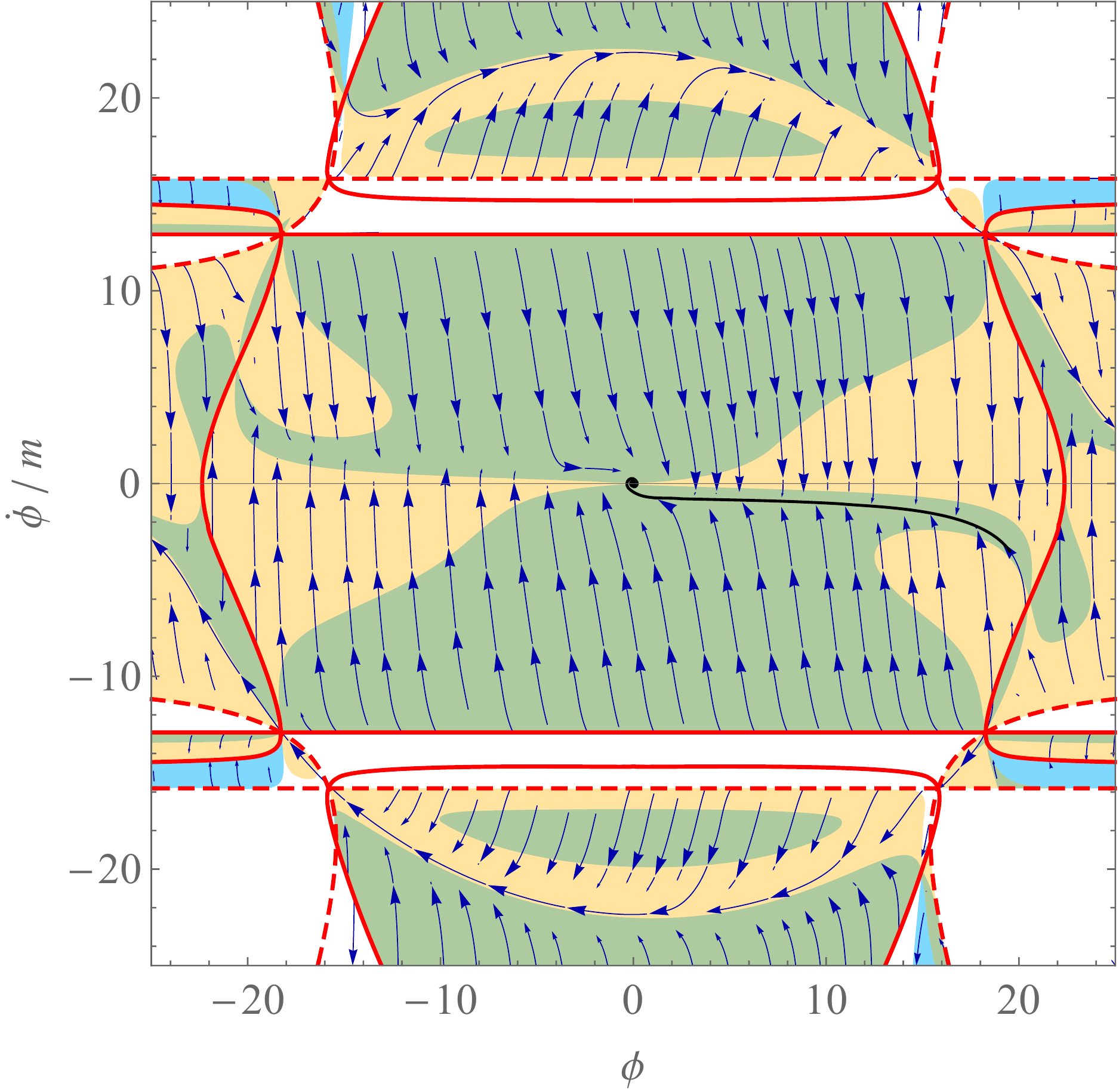}
\includegraphics[width=.45\textwidth]{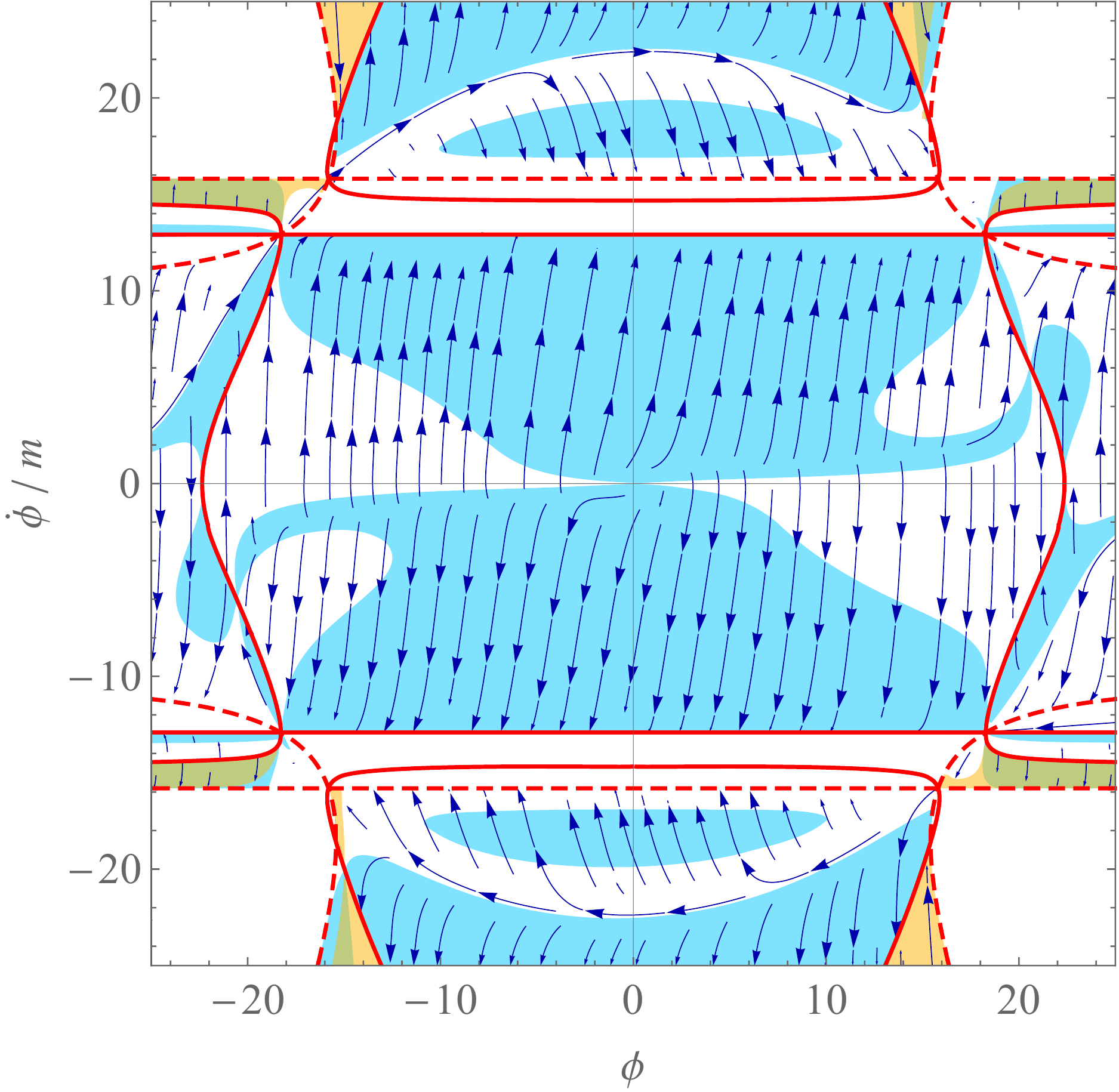}
\caption{Phase space portrait of the D-inflation model~\eqref{infmodel} with $V=m^2\phi^2/2$ and  $m^2 A_2=0.002$ and branches  $\sigma=+1$ (left), $\sigma=-1$ (right).  Shown are 
the general trajectories (blue arrows where $H^2>0$); main attractor trajectory (black)  starting from $60$ efolds before the end of inflation; 
regions with $H>0$ (yellow), $\e_H>0$ (blue) or both (green); 
curves with  $H$ singular (solid red) or changing from real to complex (dashed red).  
The attractor trajectory inevitably crosses into a region where
$\e_H<0$ while spiraling around the origin after the end of inflation.
}
\label{fig:phase-H}
\end{figure}
% ==================== Figure 1 ====================

The phase space portrait for this set of parameters is depicted in Fig.~\ref{fig:phase-H} for both branches.   Singularities in $H$ from~\eqref{red-boundary} (red solid lines) separate the phase space into disconnected 
regions with regions where $H>0$ (yellow), $\e_H>0$ (blue) or both (green) shaded.  Trajectories (blue arrows) flowing from the boundary given by the first of the conditions in
\eqref{red-boundary} (curved solid red)  with $\e_H>0$ begin at $H=+\infty$, whereas with $\e_H<0$ begin
in a collapsing phase with $H\rightarrow -\infty$ at the boundary but then bounces without a curvature singularity when $H=0$ and becomes an expanding phase $H>0$ too rapidly to be resolved in Fig.~\ref{fig:phase-H}.  
The same is true for trajectories flowing into the boundaries but with reversed signs for $H$.
These nonsingular bounces are generally accompanied by a ghost or gradient instability in the scalar or tensor sector of unitary gauge.
Trajectories flowing from the boundary given by the second of the conditions in \eqref{red-boundary} (horizontal solid red) originate from $H=+\infty$, whereas $H$ is complex on the
other side of the boundary.

From Fig.~\ref{fig:phase-H}, we see several other novel features of this model.  First, physical solutions do not exist for all possible initial phase space points:  there are regions where no real solution of $H$ 
exists on either branch.  
This occurs outside the boundaries \eqref{dashed-red-boundary} (dashed red, no trajectories), e.g.\ $\phi=\dot\phi/m=20$.

Furthermore, some trajectories in the upper and lower disconnected regions of Fig.~\ref{fig:phase-H} appear to end  at boundaries across which $H$ becomes complex by satisfying either the first (dashed red curves) or the second condition  (horizontal dashed red lines) in \eqref{dashed-red-boundary}.   Note that at these boundaries
$H$ is finite so that they do not  represent curvature singularities.  
In these cases, as mentioned below \eqref{taylor2},
the trajectories actually sharply turn so as to be tangent to the boundary at intersection.  
At intersection, the two branches become degenerate and so solutions continue on the
opposite branch, forming a parabola around this point.  In other words, trajectories staring on one branch rebound off the boundary into the opposite
branch so as to never enter the phase space region where only complex solutions exist.  

Additionally, near this rebound of the trajectories the contracting solution $H<0$ bounces to expansion $H>0$ as well.  
For the most cases, the bounce itself, $H=0$, occurs within one branch before or after hitting the dashed boundaries, whereas the branch change occurs with the rebound of the trajectories at the boundaries with finite (positive or negative) $H$, which is continuous through the rebound.  From \eqref{taylor2}, we see that there exists an exceptional case for this boundary, which is $\phi=0$, since in this case $H=0$ at the boundary and hence $H=0$ and branch change occurs at the same time. Again, this
nonsingular bounce is generally accompanied by a ghost or gradient instability in the scalar or tensor sector of unitary gauge.  
On the other hand, the condition~\eqref{c1dxmc4} for the well-definedness of unitary gauge perturbations
is itself violated around bounce solutions where the field transits a region where
$\Omega =0$ or $\pm \infty$.   In these cases, a covariant treatment or full numerical solution is required to assess perturbation
pathologies (see also \cite{Ijjas:2017pei,Dobre:2017pnt}).
For instance, on the second boundary of
\eqref{dashed-red-boundary} where $H_b$ diverges at finite $H$,
$\Omega=\pm \infty$.   
Note that at this boundary, $H_b$ and the
original higher-order equations of motion \eqref{Eq1}, \eqref{Eq2} appear 
discontinuous between the branches but when reduced to a second-order system, 
the two branches of $\ddot \phi(\phi,\dot\phi)$ join.   This property is unique to degenerate models.
Finally, there is also a novel feature that
some trajectories have the field roll up hill,
but we shall see that in general these regions are associated with ghost or gradient instability as well.

\begin{figure}[t]
\centering
\includegraphics[width=.45\textwidth]{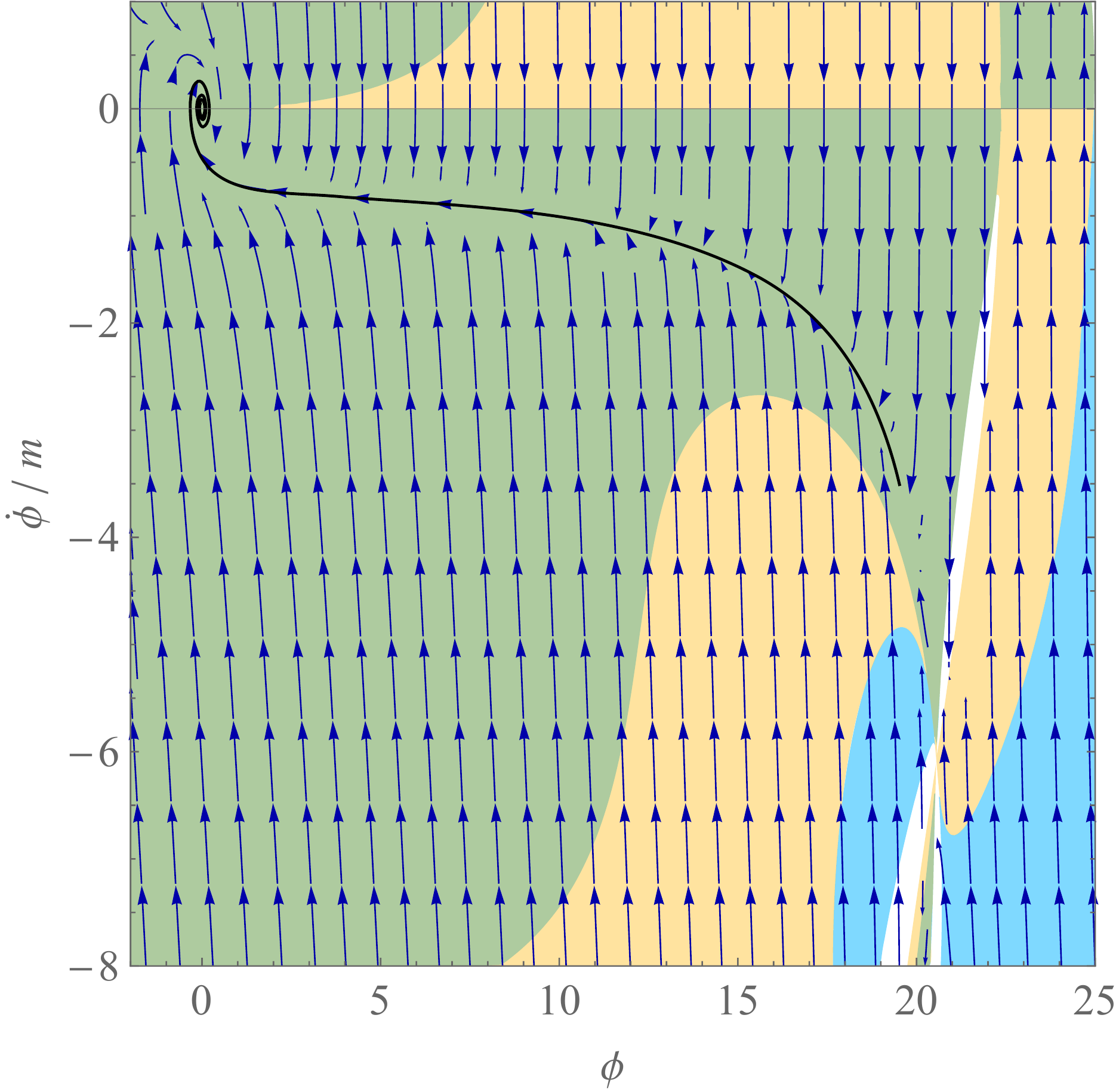}
\caption{Phase space regions for the $\sigma=+1$ branch where $b_s\e_H/c_s^2>0$  (yellow),  $c_s^2>0$ (blue), or both (green), with general trajectories (blue arrows) and attractor trajectory (black curve). 
The attractor trajectory inevitably crosses into a regime of gradient instability 
when spiraling around the origin (see also Taylor expansions \eqref{grad-inst} and Fig.~\ref{fig:ghost-gradient-const-phi}). 
}
\label{fig:ghost-gradient}
\end{figure}

On the other hand, the trajectories in the central region of Fig.~\ref{fig:phase-H} for $\sigma=+1$ are similar to the canonical ones.  Also as in GR, trajectories start or end on singularities (solid red curves), albeit here at finite field values. 
This region also exhibits an attractor solution which is visually apparent from the converging flows in
Fig.~\ref{fig:ghost-gradient}.   To isolate this trajectory we numerically integrate the reduced evolution equation $\ddot\phi=\ddot\phi(\phi,\dot\phi)$.
For the initial condition, we adopt $(\phi,\dot\phi/m) = (20.3, -6)$ at $t=0$, which is close to the intersection of the singularity and the attractor and rapidly evolves onto the attractor.
The numerical solution $\phi=\phi(t)$ on the attractor (black curve, the left panel of Fig.~\ref{fig:phase-H}) is shown for the 60 efolds before the end of inflation. Time $t$ is 
converted to efolds $\N$ by plugging the numerical solution $\phi=\phi(t)$ into the equation $H=H(\phi,\dot\phi)$ and numerically integrating it.
We place the zero point of  efolds at the end of inflation, i.e.\ $\epsilon_H=1$ at $\N=0$.
Note that $\N=0$ at $(\phi,\dot\phi/m)\approx (1.0, -0.71)$ and $\N=-60$ 
at $(\phi,\dot\phi/m)\approx (20, -3.5)$.   
During inflation on the attractor unitary gauge perturbations are well-defined since $\Omega$ given in \eqref{Omega} is finite and nonzero.

After inflation when $\N>0$, the attractor trajectory spirals around the origin and inevitably crosses
into a region of noncanonical behavior where $\epsilon_H<0$.  We shall see next
that this region is associated with gradient instabilities.

% ==================== Figure 2 ====================
% ==================== Figure 2 ====================

\subsection{Perturbations}%%%%%%%%%%%%%%%%%%%%%
\label{ssec:3C}

The central region of Fig.~\ref{fig:phase-H}, with its attractor solution on the $\sigma=+1$ branch, provides a potentially viable inflationary regime and
we therefore focus on it for the perturbation analysis.
From the EFT coefficients \eqref{C4}, \eqref{C5} and their time derivatives, we
can construct $b_s, c_s, b_t, c_t$ and their associated slow-roll parameters.
First, the tensor sector is simple.  From \eqref{btct}, $b_t=1$ and $c_t^2 = (1 - 2 \dot\phi^2 A_2)^{-1}$ and hence the stability condition  $b_t>0$ and $c_t^2>0$ for the present case is satisfied if $|\dot\phi /m|< (2 m^2 A_2)^{-1/2}\approx 16$ as it is in the central region.

The scalar sector, parametrized by  $b_s$ and $c_s$, is more complicated.   Their explicit forms are too cumbersome to provide here, but straightforward to obtain. 
In Fig.~\ref{fig:ghost-gradient}, we show the regions where  $\epsilon_H b_s/c_s^2>0$ (yellow), $c_s^2>0$ (blue), or both (green, instability free) near the attractor of the central region. Note that in our example
these are invariant for  $(\phi,\dot\phi)\rightarrow (-\phi,-\dot\phi)$ so we
only display the lower right quadrant.     The attractor itself (black line) remains in the stable region
from the end of inflation  to $\sim 60$ efolds prior, but approaching it especially from small velocities may require crossing from a region of gradient instability.

At reheating, trajectories spiral around the origin and cross
$\dot\phi=0$.  
Here the field will inevitably enter into an unstable regime as we can analytically check as
follows.
For both branches of $H_b$ and a general potential,
the Taylor expansions of $H_b/H$ and $\dot H$ around $\dot\phi=0$ are given by 
\begin{align} \label{HbdHexp}
H &= \sigma\sqrt{\f{V}{3}}\kk{ 1 - \sigma a_1 A_2 \dot\phi + \O(\dot\phi^2) }, \notag\\
H_b &= \sigma\sqrt{\f{V}{3}} \kk{1 + \f{\dot\phi^2}{4 V} + \O(\dot\phi^4) } , \notag\\ 
\dot H &= \f{A_2 V}{3} \kk{ a_1^2 + 8 \sigma a_1 \dot\phi + \O(\dot\phi^2) }, \notag\\
\dot H_b &=  - \f{A_2 V}{3} \sigma a_1 \dot\phi - \f{\dot\phi^2}{2} + \O(\dot\phi^3) ,
\end{align}
where $a_1 \equiv \sqrt{\f{3}{V}} \f{V'}{1 - 2 A_2 V}$.
Note that 
the Taylor expansion of $H_b$ does not include odd powers of $\dot\phi$ since in our model $H_b$ in \eqref{Hbsols} is a function of $\dot\phi^2$.
The leading order behavior of  $\dot H$ is a constant as $\dot\phi\to 0$ which vanishes if $A_2\to 0$.   In our model, this is a positive constant so unlike a canonical field
$\e_H<0$ as $\dot\phi\to 0$ as shown in Fig.~\ref{fig:phase-H}.

% ==================== Figure 3 ====================
\begin{figure}[t]
\centering
\includegraphics[width=.43\textwidth]{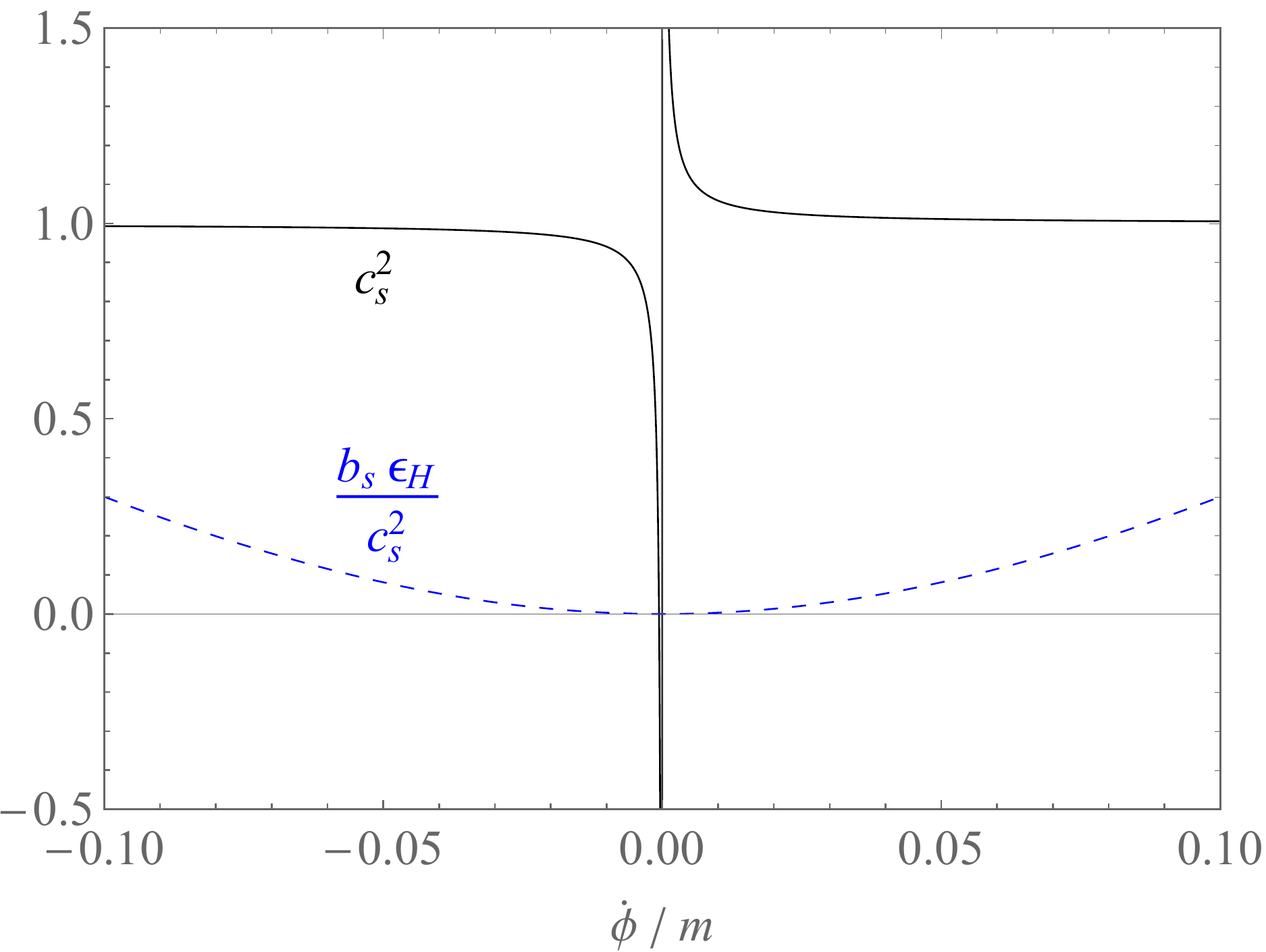}
\caption{
Behavior of $c_s^2$ (solid black)  and  $b_s \e_H/c_s^2$ (dashed blue) and as a function of $\dot\phi/m$ near the origin at fixed $\phi=-0.3$.  At $\dot\phi=0$, $c_s^2 \rightarrow \pm \infty$ when approached from alternate sides.  Similar behaviors can be observed for $\phi=0.3$.  There exists gradient instability near the origin, while no ghost instability.
}
\label{fig:ghost-gradient-const-phi}
\end{figure}
% ==================== Figure 3 ====================

Using \eqref{HbdHexp}, we can expand the sound speed and normalization as
\begin{align}
\lim_{\dot\phi\to 0} c_s^2 &= \f{(H - H_b)(4 H + H_b) - \dot H_b}{6 (H - H_b)^2 },\notag\\
&= -\f{2 \sigma}{3 a_1 A_2 \dot\phi} \kk{1+\O(\dot\phi)} , \nonumber\\
\lim_{\dot\phi\to 0} \frac{b_s \e_H}{c_s^2}  
&= \f{6 (H - H_b)^2}{H_b^2 }, \notag\\
&= 6 a_1^2 A_2^2 \dot\phi^2 \kk{1+\O(\dot\phi)}.
\end{align}
The normalization $b_s \e_H/c_s^2$ remains positive but approaches zero as $\dot\phi \rightarrow 0$, while
$c_s^2$ diverges in amplitude.
Notice that this divergence occurs even as $A_2\to 0$, despite the fact that $c_s^2=1$ for
$A_2=0$, which indicates a discontinuous limit. 
For the potential $V(\phi)=m^2\phi^2/2$, 
\be \label{grad-inst} \lim_{\dot\phi\to 0} c_s^2 
= - \sqrt{\f{2}{3}} \f{m}{\dot\phi} \f{\sigma \phi}{|\phi|}  \f{ 1 - A_2 m^2 \phi^2}{3  m^2 A_2 } \left[ 1+ \O(\dot\phi)\right] . \ee  
Therefore, for both branches, near the origin where $1 - A_2 m^2 \phi^2>0$ is satisfied,
the sign of $c_s^2$ is determined by $\phi /\dot\phi$.
Hence, for $\sigma =+1$ branch, $c_s^2 \rightarrow -\infty$ at the first and third quadrants and $c_s^2 \rightarrow +\infty$ in
the second and fourth quadrants, where the attractor originates.   
In the former, $c_s^2<0$ occurs only in a small
neighborhood around $\dot\phi=0$ as shown in Fig.~\ref{fig:ghost-gradient-const-phi}.

Furthermore, from \eqref{Omega},
\begin{align} \label{Omega-y-0}
\lim_{\dot\phi\to 0} \f{\Omega}{6a^3} &= -H_b 
= -\sigma\sqrt{\f{V}{3}} \kk{1 + \O(\dot\phi^2) },
\end{align}
and hence exactly at the origin $(\phi,\dot\phi)=(0,0)$ of our model, the condition~\eqref{c1dxmc4} is violated with $\Omega=0$ and the unitary gauge becomes ill-defined.

To avoid gradient instability and unitary gauge being ill-defined,
we can relax the assumption that $A_2={\rm const}$ and
choose $A_2(\phi,X) \rightarrow 0$ at the origin.  Provided this occurs only  for $|\phi|, |\dot \phi/m| \lesssim 1$, the  dynamics of perturbations during inflation  will not be affected.

% ==================== Figure 4 ====================
\begin{figure}[t]
\centering
\includegraphics[width=.45\textwidth]{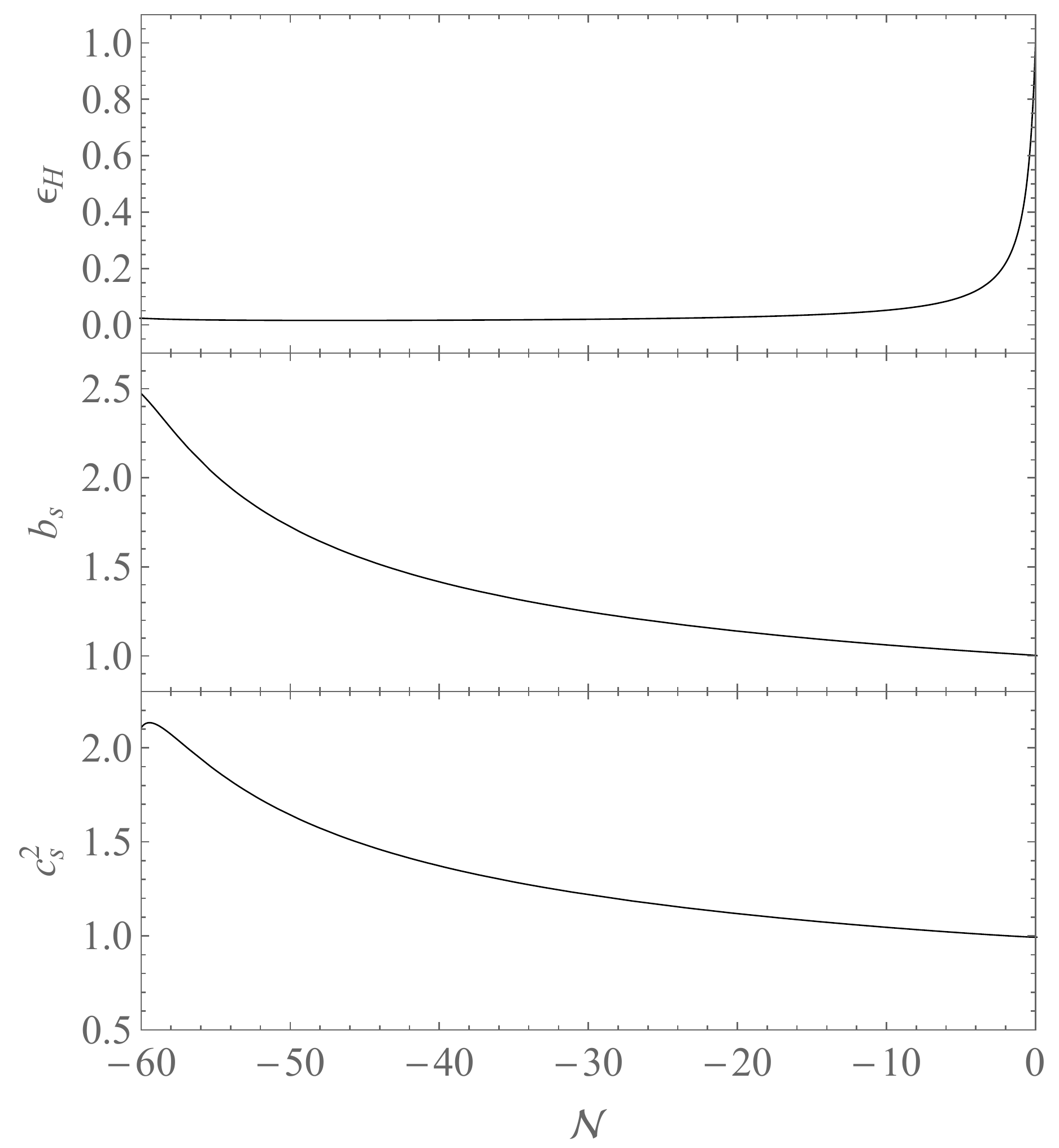}
\includegraphics[width=.45\textwidth]{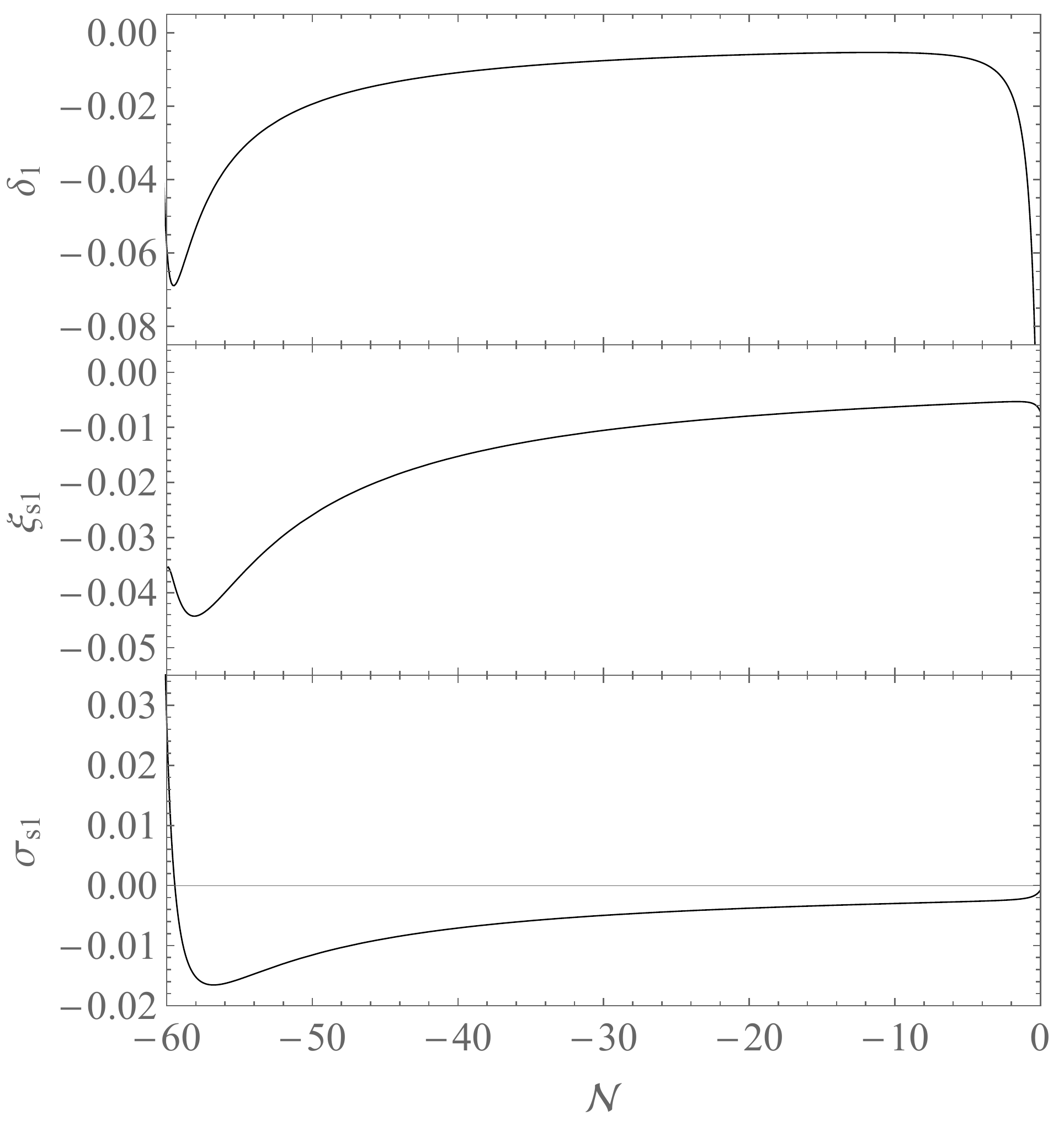}
\caption{Variation of $\epsilon_H, b_s, c_s^2$, and corresponding slow-roll parameters $\delta_{1}, \xi_{s1}, \sigma_{s1}$ along the attractor.
}
\label{fig:EFTcoeff-att-N}
\end{figure}
% ==================== Figure 4 ====================

Finally we can examine the evolution of $b_s$ and $c_s^2$ along the attractor during inflation.
In Fig.~\ref{fig:EFTcoeff-att-N} we show variation of $\e_H, b_s, c_s^2$, and corresponding slow-roll parameters~\cite{Motohashi:2017gqb}
\be
\d_1 \equiv \f{1}{2}\f{d\ln \e_H}{d\N} - \e_H , \quad
\xi_{s1} \equiv \f{d\ln b_s}{d\N} , \quad 
\sigma_{s1} \equiv \f{d\ln c_s}{d\N} , 
\ee
along the attractor. 
They are defined based on the quadratic action~\eqref{S2s} for $\tilde\zeta$, but as we mentioned above in our model $\tilde\zeta- \zeta$ evolves to zero so $\ln \Delta_{\tilde\zeta}^2 = \ln \Delta_{\zeta}^2$ after inflation. 
Notice that while all remain perturbative, $\sigma_{s1}$ in particular can become moderately large around $\N\sim -60$ and moreover evolves on the several
efold time scale.

Such cases can be treated in the optimized slow-roll (OSR) formalism~\cite{Motohashi:2015hpa,Motohashi:2017gqb}, where
the slow-roll (SR) result for 
the curvature power spectrum after inflation when $\tilde\zeta=\zeta$ 
\begin{align} \label{lnD2SR}
\ln \Delta_\zeta^2 \Big|_{\rm SR} = \ln\mk{\f{H^2}{8 \pi^2 b_s c_s \e_H}} ,
\end{align}
is corrected by the slow-roll parameters
\begin{align} \label{lnD2OSR}
\ln \Delta_\zeta^2  \approx  \ln \Delta_\zeta^2 \Big|_{\rm SR}
 - \f{10}{3}\e_H - \f{2}{3}\d_1 - \f{7}{3}\sigma_{s1} - \f{1}{3}\xi_{s1}  ,
\end{align}
and evaluated at freeze-out where $k\int_{\N}^0 d\N c_s /a H \approx e^{1.06}$, contrary to  $k\int_{\N}^0 d\N c_s /a H=1$ in the SR approximation.

\begin{figure}[t]
\centering
\includegraphics[width=.45\textwidth]{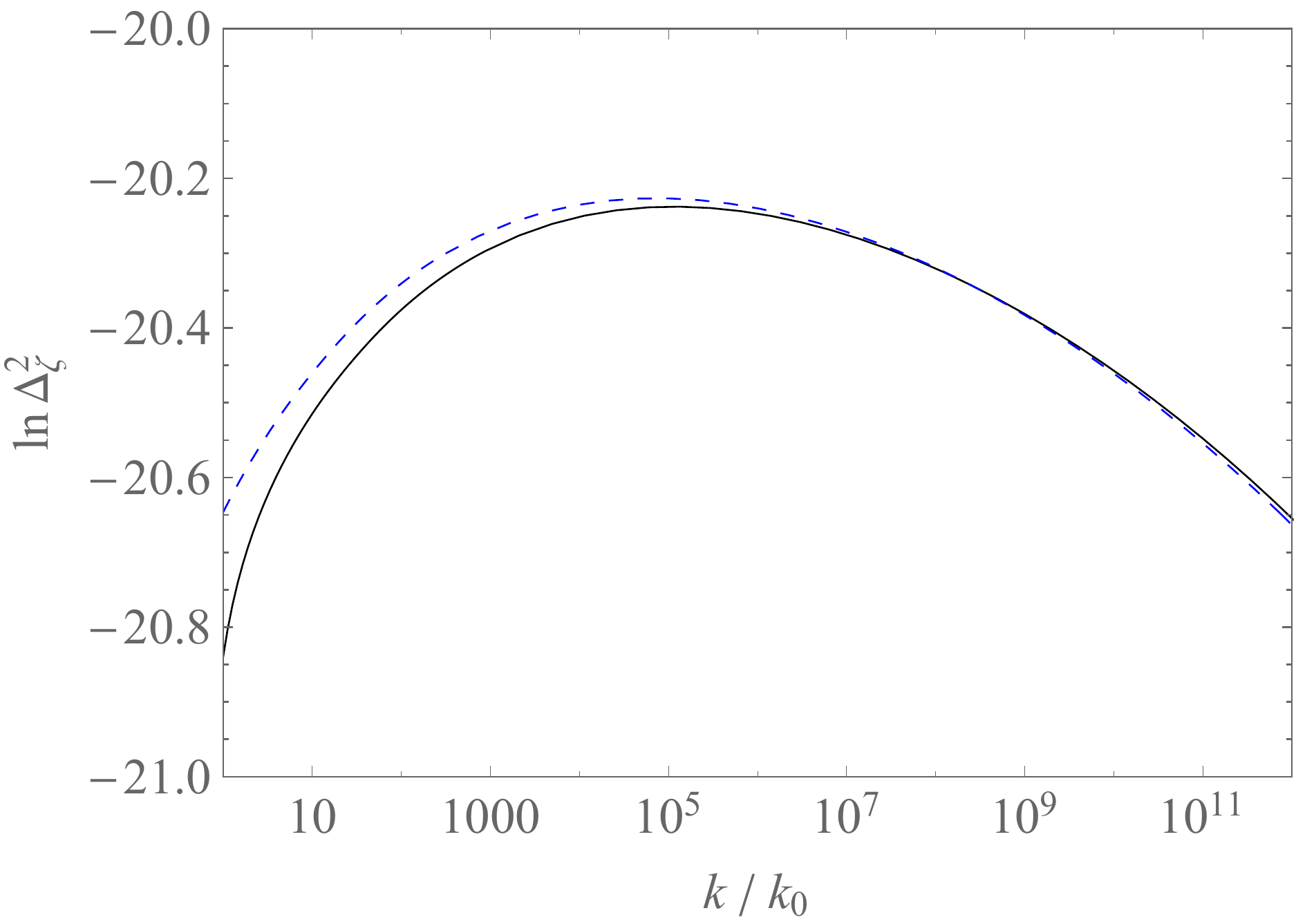}
\caption{The curvature power spectrum $\ln \Delta_\zeta^2$ evaluated under the improved
OSR approximation \eqref{lnD2OSR} (solid black) compared with the SR approximation (dashed blue).   $k_0$ represents the mode that freezes out in the OSR approximation at
$\N =-60$.  
}
\label{fig:lnD2}
\end{figure}

These approximations are compared in Fig.~\ref{fig:lnD2} for the same $k$, where $k_0$ is the mode that freezes out at $\N=-60$ in OSR.
Here we additionally choose $m=10^{-5}$ to fix the normalization of $H$ in Planck units and hence that of $\Delta_\zeta^2$ to be roughly compatible with observations.
Notice that there is a significant running of the tilt pivoting around $k/k_0 \sim 10^4$ or $\N\sim -50$ despite being far from the 
end of inflation and containing no  features in the potential there. 
In this region, the OSR and SR results differ in shape and OSR itself breaks
down as an approximation for some $\N < -60$ where the corrections become order unity.
The OSR approximation thus extends the regime of validity for the calculation into the range  $-60 \lesssim \N \lesssim -50$, which is relevant for the CMB, and is useful
in observationally constraining D-inflation.
We leave such a study and the construction of an observationally viable model to a future work.

% ==================== Figure 5 ====================
% ==================== Figure 5 ====================

\bigskip

\section{Conclusion}%%%%%%%%%%%%%%%%%%%%%%%%%%%%%%%%%%%%%%%%%
\label{sec:con}

In this work, we developed the EFT of inflation for a general Lagrangian constructed from ADM variables, which encompasses the most general interactions with up to second derivatives of a scalar field whose background spontaneously breaks temporal 
diffeomorphism symmetry. 
The Ostrogradsky ghost usually implied by such higher-order terms is eliminated by degeneracy conditions, leading to degenerate higher-order (or D-)inflation.  We identify 8 types of degeneracy conditions, one of which corresponds to  known DHOST models.   
For the other cases, which include curvature couplings not considered in DHOST, we provide necessary conditions for a covariant scalar-tensor theory based on the dispersion relation of the quadratic action  and leave
a full assessment of their viability to future work.  

Higher-order theories imply equations of motions that are  higher than second order in the scalar field and typically lead to an ill-posed Cauchy problem.   
The degeneracy conditions, which involve the metric as well, restores a well-defined forwards or backwards evolution from initial field and field derivative
data  on a Cauchy surface but with novel features.   

We illustrate these features
with an explicit example of D-inflation.  
First, not all field configurations lead 
to physical solutions for the metric as illustrated by values where all solutions for
the Hubble parameter become complex even for positive potentials and timelike field gradients.   
Second, evolution is only uniquely defined up to a branch choice since the same field configurations lead to distinct
expansion histories that are not related by time reversal as they would be in GR.  This feature is present in Horndeski theory as well.  Third, trajectories can sharply turn to avoid
phase space regions where real solutions fail to exist leading to highly complicated phase space portraits
where contraction can turn to expansion without encountering a curvature singularity.    These bouncing solutions generally traverse regions of
ghost or gradient instabilities in unitary gauge but also cross coordinate singularities in defining its metric perturbations
(see also~\cite{Ijjas:2017pei,Dobre:2017pnt}).
Finally, perturbations can go unstable even in the limit that the additional
degenerate terms in the Lagrangian are infinitesimal.   In our example this occurs
for curvature perturbations in the simplest model
of constant, but arbitrarily small, higher-order coefficients
during reheating when the inflaton oscillates around the minimum of its potential. 

Our D-inflation model also has novel phenomenology.   
While the model possesses an attractor which leads to nearly scale invariant fluctuations across a sufficient number of
efolds of inflation, it also can produce substantial running of the tilt on CMB scales despite having
no features in the potential there and being far from the end of inflation.   In this case, EFT coefficients vary on the several
efold timescale and require an approach that goes beyond the usual slow-roll formalism.  
We show that corrections captured by the optimized slow-roll approach extends the
validity of predictions into the large running regime of interest and should be useful
in observational tests of the D-inflation scenario.

\acknowledgments%%%%%%%%%%%%%%%%%%%%%%%%%%%%%%%%%%%%%%%%%
We thank  Marco Crisostomi, Jose Maria Ezquiaga, Kazuya Koyama, and Sam Passaglia for useful comments.  H.M.\ was supported by Japan Society for the Promotion of Science (JSPS) Grants-in-Aid for Scientific Research (KAKENHI) No.\ JP17H06359 and No.\ JP18K13565.
W.H.\ was supported by U.S.~Dept.~of Energy Contract No.\ DE-FG02-13ER41958 and the Simons Foundation.

\appendix

\section{Degeneracy Conditions}%%%%%%%%%%%%%%%%%%%%%%%%%%%%%%%%%%%%%%%%%
\label{app:deg}

We can determine the necessary conditions for degeneracy by examining the high $k$ or high frequency limit of the
quadratic Lagrangian.  We can then find the number of propagating modes and their dispersion relation by
assuming 
solutions of the form $u(x,t) = u(k) e^{i(\omega t+ k x)}$ where  $u=(\delta N, \zeta, \psi)^T$ \cite{Langlois:2017mxy}.

In the limit, we can  neglect evolution on the Hubble time scale of the background and the EFT coefficients up to corrections of order $(k/aH)^2$, which as we detail below is sufficient to establish degeneracy conditions for most solutions and easily supplemented in the remaining ones.
The quadratic Lagrangian~\eqref{L2s-c} for scalars can be then written as 
\be \L_2 = \f{1}{2}u^\dagger K u ,\ee
with the kinetic matrix
\be
K\equiv \bem
c_9+\omega^2 c_3 + c_{10} \f{k^2}{a^2} 
& \omega^2c_2 + i\omega \mk{ c_4 + c_5\f{k^2}{a^2} } + c_8 \f{k^2}{a^2} 
& -i\omega \frac{c_2}{3} \f{k^2}{a^2} + \frac{c_4}{3} \f{k^2}{a^2} \\
\omega^2c_2 - i\omega \mk{ c_4 + c_5\f{k^2}{a^2} } + c_8 \f{k^2}{a^2} 
& \omega^2c_1 + c_6 \f{k^2}{a^2} + c_7 \f{k^4}{a^4} 
& -i\omega \frac{c_1}{3} \f{k^2}{a^2} + c_{12} \f{k^4}{a^4} \\
i\omega \frac{c_2}{3} \f{k^2}{a^2} + \frac{c_4}{3} \f{k^2}{a^2} 
& i\omega \frac{c_1}{3} \f{k^2}{a^2} + c_{12} \f{k^4}{a^4} 
& c_{11} \f{k^4}{a^4}
\eem .
\ee
For nontrivial solutions of the equation of motion $K u=0$ to exist, 
we require $\det K=0$, which can be written as
\be \label{disp} 
f_1 \omega^4 + \ck{ f_2 \mk{\f{k}{a}}^4 + f_3 \mk{\f{k}{a}}^2 + f_4 } \omega^2 + f_5 \mk{\f{k}{a}}^6 + f_6 \mk{\f{k}{a}}^4 + f_7 \mk{\f{k}{a}}^2 = 0 , \ee
where
\begin{align} \label{fcoef}
f_1 
&\equiv -\mk{ \frac{c_1}{9}- c_{11}} (c_1 c_3 - c_2^2),
\notag\\
f_2 
&\equiv c_3 ( c_7 c_{11} - c_{12}^2) - \f{1}{9} c_2^2 c_7 - \f{2}{3} c_2 c_5 c_{12} - c_5^2 c_{11},\notag\\
f_3 
&\equiv \f{c_6}{9} (c_1 c_3 - c_2^2) - \mk{ \f{c_1}{9} - c_{11} } (c_1 c_{10} - 2 c_2 c_8 + c_3 c_6 - 2 c_4 c_5), \notag\\
f_4 
&\equiv\left(  \frac{c_1}{9}- c_{11} \right) ( c_4^2 - c_1 c_9 ),\notag\\
f_5 &\equiv c_{10} (c_{7} c_{11} - c_{12}^2) ,\notag\\
f_6 
&\equiv c_9 (c_7 c_{11} - c_{12}^2) 
+  c_6 c_{10} c_{11} - \f{1}{9} c_4^2 c_7 + \f{2}{3} c_4 c_8 c_{12} - c_8^2 c_{11}, \notag\\
f_7 &\equiv c_{6} \left(c_{9} c_{11} -\frac{c_4^2}{9}\right) .
\end{align}
In general, this is a fourth order system for $\omega$ representing two propagating modes.
To remove the second propagating mode in unitary gauge, 
we demand $f_1=0$,
for which there are several possibilities: 
\begin{enumerate}
\item 
$c_3 = c_2^2/c_1$ or equivalently
$\displaystyle \C_{\beta\beta} = 3 \C_{\beta K}^2/(3 \C_{KK} + \tilde \C_{KK})$.  The kinetic terms organize into
a single term for $\tilde \zeta = \zeta + (c_2/c_1)\delta N$, which is the propagating degree of freedom.
\item
$c_1 = c_2=0$ or equivalently
$ 3\C_{KK}+\tilde\C_{KK}=\C_{\beta K}=0 $.  The kinetic term for $\zeta$ vanishes and $\delta N$ is the propagating
degree of freedom.
\item 
$c_{11} = c_1/9$ or equivalently
$\tilde\C_{KK}=0$.  The constraint equation for $\psi$ eliminates the kinetic term for $\zeta$ and $\delta N$ is again the
propagating degree of freedom.
\end{enumerate}
Below we shall consider each case in Appendix~\ref{app:case1}, \ref{app:case2}, \ref{app:case3}, respectively.

Furthermore, retaining a higher order in spatial derivatives (or $k$) compared
with temporal derivatives (or $\omega$) in a fully covariant theory
corresponds to the reappearance of the second mode when changing the gauge. 
Therefore to find covariant degeneracy conditions, we seek solutions of Eq.~(\ref{disp}) that
correspond to a normal dispersion relation $\omega^2 = c_s^2 k^2$.   
The possible cases are
\begin{enumerate}
\item[a.] $f_4f_7\ne 0$, others~$=0$~~$\Rightarrow$~~$f_4\omega^2+f_7 (k/a)^2=0$.
\item[b.] $f_3f_6\ne 0$, others~$=0$~~$\Rightarrow$~~$f_3\omega^2+f_6 (k/a)^2=0$.
\item[c.] $f_2f_5\ne 0$, others~$=0$~~$\Rightarrow$~~$f_2\omega^2+f_5 (k/a)^2=0$.
\end{enumerate}

There are several caveats regarding this technique that need to be borne in mind.  Since we neglect Hubble scale
evolution, we work in the limit $\omega \gg H$ and since the $c_i$ coefficients generically carry a mass dimension $M_i^{n_i}$
for some $n_i$ and can vary on the Hubble time scale we assume $\dot c_i/c_i \sim H \ll M_i$ as well. Whereas the former condition corresponds to $c_s k/aH \gg 1$ for a linear 
dispersion relation, the latter need not necessarily be small in practice.   Since we are mainly interested in this technique for
deriving degeneracy conditions  and the form of the dispersion relation, rather than the exact coefficients in the dispersion relation,
this technique suffices.   The exception is when $(c_s k/aH)^{-2}$ corrections spoil the form of the dispersion relation at
$c_s k/aH\gg 1$.    This can occur in the ``a" case through corrections to $f_2$ and $f_5$ which can then dominate over the terms  from $f_4$ and  $f_7$ which form
the desired linear dispersion relation.    For the ``b" case these corrections can change the
coefficients but not the leading order form and for the ``c" case, the corrections from the other terms are entirely
negligible for $c_s k/aH \gg 1$.   We therefore further check for supplemental degeneracy conditions in the  ``a"  or $f_4 f_7 \ne 0$ case.
Note that since the coefficients in the dispersion relation can also change in the ``a" and ``b" cases from those
given by this static technique, the full quadratic Lagrangian should be used to check for ghost and gradient instabilities in 
those cases.

We now consider the various kinetic structures  $1,2,3$ and their degeneracy conditions under $a,b,c$ respectively.
We treat ``1a" in more detail as it serves both as an example of the technique and includes the known DHOST models.

\subsection{$c_3=c_2^2 /c_1$ case}%%%%%%%%%%%%%%%%%%%%%
\label{app:case1}

Plugging $c_3=c_2^2 /c_1$ to \eqref{fcoef} we have $f_1=0$ and reduced forms for $f_{2\ldots 7}$ which imply
 there is a single propagating degree of freedom 
\be \tilde \zeta = \zeta + \f{c_2}{c_1}\delta N \ee
in unitary gauge.  
The degeneracy classes $a,b,c$ where this degree of freedom obeys a linear dispersion relation for $c_s k/a H\gg 1$
are defined by the pair of $f$ coefficients that remain nonzero.    
Therefore in each case, we have 5 degeneracy conditions between 
the various EFT coefficients represented by $c_i$ in the static limit.
Case 1a can have  supplementary degeneracy conditions beyond the static limit as discussed above.

\begin{itemize}

\item[a.] First, let us consider the case $f_4 f_7\ne 0$ and all other $f$'s zero.    
Requiring first that $f_5=0$ leads to two branches
\be \label{f5branch}
\quad c_{12}^2=c_7 c_{11} \quad {\rm or} \quad
\quad c_{10}=0, 
\ee
For each case, $f_2=f_3=f_6=0$ should be satisfied.  
While in general $f_3=0$ has two branches, 
since $f_4\ne 0$ implies $c_{11} \ne c_1/9$, only one solution remains
\be \label{1af3} c_1^2 c_{10} = 2 c_1 (c_2 c_8 + c_4 c_5) - c_2^2 c_6. \ee
Also, $f_2=0$ yields
\be \label{1af2} c_5^2 c_{11} + \f{1}{9} c_2^2 c_7 + \f{2}{3} c_2 c_5 c_{12} = \f{c_2^2}{c_1} (c_7 c_{11} - c_{12}^2), \ee
and $f_6=0$ yields
\be \label{1af6} c_9 ( c_7 c_{11} - c_{12}^2) + c_6 c_{10} c_{11} - c_8^2 c_{11} - \f{1}{9} c_4^2 c_7 + \f{2}{3} c_4 c_8 c_{12}=0.
\ee
These are the degeneracy conditions for the static, high $k$ limit.

More generally, this ``1a'' case is subject to corrections which require supplementary 
degeneracy conditions as described above.   
Using only the first condition $c_3=c_2^2/c_1$,
the quadratic Lagrangian~\eqref{L2s-c} reduces to 
\begin{align}
\L_2 &= \f{1}{2}c_1 \dot{\tilde\zeta}^2 + \mk{c_4-c_1\dot x + c_5 \f{k^2}{a^2}} \dot{\tilde\zeta} \delta N  
+ \f{1}{2}\mk{c_6 +c_7\f{k^2}{a^2}}\f{k^2}{a^2} \tilde\zeta^2 
+ \mk{ c_8-c_6x -c_7x\f{k^2}{a^2}} \f{k^2}{a^2} \tilde\zeta \delta N 
\notag\\ &~~~
+ \f{1}{2}\mk{\tilde c_9 + \tilde c_{10} \f{k^2}{a^2} + c_7x^2\f{k^4}{a^4}} \delta N^2 
+ \f{1}{2}c_{11} \f{k^4}{a^4} \psi^2 
+ \f{k^2}{a^2} \psi \kk{ \f{c_1}{3} \dot{\tilde\zeta} + \f{c_4-c_1\dot x}{3} \delta N 
+ c_{12} \f{k^2}{a^2} (\tilde\zeta-x\delta N)},   
\label{L21a}
\end{align}
where $x\equiv c_2/c_1$ and 
\begin{align}
\tilde c_9 &\equiv c_9 + \dot c_4 x + c_1 \dot x^2 - c_4 \dot x,\notag\\
\tilde c_{10} &\equiv c_{10} + (\dot c_5 - 2 H c_5 -2 c_8 ) x + c_6 x^2 - c_5 \dot x .
\end{align}
Potentially problematic terms are those where the fields have  time derivatives and the coefficients carry  additional 
factors of $k^2/a^2$.   In the static limit, these terms are arranged to cancel, but beyond the static limit the
time variation of the coefficients breaks this degeneracy relation and changes the dispersion relation even for $c_s k/aH\gg 1$.
The only term of this form is $c_5 (k/a)^2 \dot{\tilde\zeta}\delta N$.   Therefore $c_5=0$ is sufficient as 
a supplemental degeneracy condition to ensure a linear dispersion relation for the the single propagating
degree of freedom $\tilde \zeta$.   This condition may be generalized to nonvanishing $c_5$ but would then
involve tuning between $c_5, a$ and the other $c_i$ coefficients.   Due to the appearance of the scale factor $a$
in the generalized degeneracy condition
this tuning is unlikely to be preserved in a fully covariant scalar-tensor theory.    We therefore take $c_5=0$ and the complete set of degeneracy
conditions for case 1a has two branches
\begin{eqnarray}
i) \quad && c_3=\f{c_2^2}{c_1},  \quad c_{12}^2=c_7 c_{11} , \quad {c_1^2} c_{10} = {2 c_1 c_2 c_8  - c_2^2 c_6 }, \quad  c_2^2 c_7  = \f{9c_2^2}{c_1} (c_7 c_{11} - c_{12}^2),\nonumber\\
\quad&& 
 c_9 ( c_7 c_{11} - c_{12}^2) + c_6 c_{10} c_{11} - c_8^2 c_{11} - \f{1}{9} c_4^2 c_7 + \f{2}{3} c_4 c_8 c_{12}=0 , \quad c_5=0, 
 \label{1ai} \\
 ii) \quad && c_3=\f{c_2^2}{c_1}, \quad  c_{10}=0, \quad {2 c_1 c_2 c_8  - c_2^2 c_6 }=0,\quad
c_2^2 c_7 = \f{9c_2^2}{c_1} (c_7 c_{11} - c_{12}^2), \nonumber\\
\quad && c_9 ( c_7 c_{11} - c_{12}^2) + c_6 c_{10} c_{11} - c_8^2 c_{11} - \f{1}{9} c_4^2 c_7 + \f{2}{3} c_4 c_8 c_{12}=0 ,\quad c_5=0.
\label{1aii}
\end{eqnarray} 
With this complete set of degeneracy conditions, 
one can explicitly verify that the Euler-Lagrange equations that result from Eq.~(\ref{L21a}) describe a single propagating 
degree of freedom $\tilde \zeta$ with a linear dispersion relation at $c_s k/aH \gg 1$.

While the degeneracy conditions in Eq.~(\ref{1ai}) or (\ref{1aii}) are complete, they allow for a variety of
ways that the EFT coefficients can satisfy them.   To make the connection with DHOST models, we can further
examine these explicit solutions.   
For instance, for $1a$-$i$ we can have 
\be \label{D1}
c_3=\f{c_2^2}{c_1}, \quad 
c_5=c_7=c_{11}=c_{12}=0, \quad 
c_{10} = \frac{2 c_1 c_2 c_8  - c_2^2 c_6 }{c_1^2}, \ee
or
\be \label{D2}
c_3=\f{c_2^2}{c_1}, \quad  c_5=c_7=c_{12}=0, \quad
c_8 = \f{c_2 c_6}{c_1}, \quad
c_{10} = \f{c_2^2 c_6}{c_1^2}, \quad
\ee
and for $1a$-$ii$ 
\begin{align} \label{D3}
c_3 &=\f{c_2^2}{c_1}, \quad
c_5=c_{10} = 0, \quad
c_6 = \f{2 c_1  c_8}{c_2}, \quad 
c_{11} = \f{c_1}{9} + \f {c_{12}^2}{  c_7 }, \quad 
c_9 = \f{ (c_4 c_7 - 3 c_8 c_{12})^2+   c_1  c_7 c_8^2 }{c_1  c_7^2} .
\end{align}
Models where $c_{11}=0$ on the \eqref{D2} branch are also members of the \eqref{D1} branch.
On the other hand the conditions $c_8 = c_2 c_6/c_1$, $\tilde c_{10}=0$ ($c_{10}= c_2^2 c_6/c_1^2$, $c_5=0$)
must be satisfied for any model on the \eqref{D2} branch, including
those that are part of the \eqref{D1} branch.   As pointed out by  Ref.~\cite{Langlois:2017mxy}, this presents a problem if one wants to recover
Newtonian gravity for nonrelativistic matter.   Since these conditions and $c_7=0$ zero out the $k^2 \tilde \zeta \delta N$ 
and $k^2\delta N^2$ terms in (\ref{L21a}), the Euler-Lagrange
equation for $\delta N$ which usually provides a source to the Poisson equation through the matter density is absent on this branch.    Instead the $k^2\tilde \zeta$
term in its equation of motion comes from its own Euler-Lagrange equation and has a source in matter pressure.   
For this reason, in the main text we focus on the \eqref{D1} branch.
This branch also includes the 2N-I/Ia class of DHOST models \cite{Langlois:2017mxy}.

The case 1$a$-$i$ with \eqref{D1} or \eqref{D2} corresponds to DHOST class I or II, respectively, 
and the latter was known to suffer from gradient instability.  
On the other hand, the case 1$a$-$ii$ with \eqref{D3} is not included in DHOST theories, as it requires $c_7 \ne 0$, namely $8\C_{RR}+3\tilde\C_{RR}\ne 0$, 
which can originate from the existence of the quadratic curvature terms in the covariant Lagrangian.

\item[b.] Next, we consider $f_3f_6\ne 0$ and all other $f$'s zero.  By following the same procedure
as in case 1a, we obtain the following  four sets of degeneracy conditions
\begin{align} 
&i) &
  &c_3=\f{c_2^2}{c_1}, & 
  &c_{12}^2=c_7 c_{11} , & 
  &c_4=c_9=0, &
  &9c_5^2 c_{11} + 6 c_2 c_5 c_{12} + c_2^2 c_7 = 0, \notag\\
&ii)  & 
  &c_3=\f{c_2^2}{c_1}, & 
  &c_{12}^2=c_7 c_{11} , & 
  &c_6=0, \,\, c_9 = \frac{c_4^2}{c_1}, & 
  &9c_5^2 c_{11} + 6 c_2 c_5 c_{12} + c_2^2 c_7 = 0, \notag\\
&iii) & 
  &c_3=\f{c_2^2}{c_1}, & 
  &c_{10}=0, & 
  &c_4=c_9=0, &
  &9c_5^2 c_{11} + 6 c_2 c_5 c_{12} + c_2^2 c_7 = \f{9c_2^2}{c_1} (c_7 c_{11} - c_{12}^2), \notag\\
&iv) & 
  &c_3=\f{c_2^2}{c_1}, & 
  &c_{10}=0, & 
  &c_6=0, \,\, c_9 = \frac{c_4^2}{c_1}, & 
  &9c_5^2 c_{11} + 6 c_2 c_5 c_{12} + c_2^2 c_7 = \f{9c_2^2}{c_1} (c_7 c_{11} - c_{12}^2).
\end{align}
In this case, corrections beyond the static approximation can change the coefficients of the
dispersion relation at $c_s k/aH\gg1$ but not the form and so these provide the complete 
conditions.

Note that the above branch is not included in DHOST theories.  
While \eqref{D1} and \eqref{D2} should satisfy $c_3=c_2^2/c_1$, $c_5=c_7=c_{12}=0$,  $c_{10} = (2 c_1 c_2 c_8 - c_2^2 c_6)/c_1^2$, the above branch satisfies the degeneracy without requiring $c_5,c_7,c_{12}$ to be vanishing.  
By definition $c_7$ and $c_{12}$ are nonvanishing in the presence of quadratic curvature terms, whereas $c_5$ is nonvanishing if Lagrangian includes terms such as $\mk{ {}^{(4)}R+\Box\phi }^2$.
Also, $i$) and $ii$) does not require any condition on $c_{10}$, and $iii$) and $iv$) requires a different condition $c_{10}=0$ which can be satisfied with $2 c_1 c_2 c_8  - c_2^2 c_6 \ne 0$ as $c_8$ does not appear in the degeneracy conditions and hence is a free parameter.

\item[c.] Finally, we consider $f_2f_5\ne 0$ and all other $f$'s zero.
This leads to six possible cases
\begin{align} 
&i)&
  &c_3=\f{c_2^2}{c_1}, &  
  &c_4=c_9=0, &
  &c_{11}=\f{c_1}{9}, \hspace{22.5mm}
  c_6 c_{10} = c_8^2 , &\notag\\
&ii)&
  &c_3=\f{c_2^2}{c_1}, &  
  &c_4=c_9=0, &
  &c_6 = \f{c_1 c_8}{c_2}, \hspace{20.5mm} c_{10} = \f{c_2 c_8}{c_1}  , &\notag\\
&iii)&
  &c_3=\f{c_2^2}{c_1}, &  
  &c_6=0, \, c_9 = \frac{c_4^2}{c_1}, &
  &c_{11}=\f{c_1}{9}, \hspace{22.5mm}
  c_8 = \f{3 c_4 c_{12}}{c_1}, &\notag\\
&iv)&
  &c_3=\f{c_2^2}{c_1}, &  
  &c_6=0, \,c_9 = \frac{c_4^2}{c_1}, &
  &c_{10} = \frac{2 (c_2 c_8 + c_4 c_5) }{c_1}, \quad  c_8^2c_{11} =
  \f{c_4^2}{c_1} ( c_7 c_{11} - c_{12}^2) + c_4 \mk{\f{2}{3} c_8 c_{12} - \f{1}{9} c_4 c_7   } , &\notag\\
&v)&
  &c_3=\f{c_2^2}{c_1}, &  
  &c_6=0,\, c_{11}=\f{c_1}{9}, &
  & c_{11}c_8^2=c_9 \mk{ \f{1}{9}c_1 c_7 - c_{12}^2 }   + c_4 \mk{  \f{2}{3} c_8 c_{12}- \f{1}{9} c_4 c_7  } = 0, &\notag\\
&vi)&
  &c_3=\f{c_2^2}{c_1}, &  
  &c_9 = \frac{c_4^2}{c_1}, \,  c_{11}=\f{c_1}{9}, &
  &c_6 c_{10} = \f{(c_1 c_8 - 3 c_4 c_{12})^2}{c_1^2} . &
\end{align}

Again, note that this branch is not included in DHOST theories as the DHOST conditions $c_3=c_2^2/c_1$, $c_5=c_7=c_{12}=0$,  $c_{10} = (2 c_1 c_2 c_8 - c_2^2 c_6)/c_1^2$ are not satisfied in general.
Clearly, the conditions $i$), $ii$) do not include $c_5$, $c_7$, $c_{12}$; 
the condition $iii$) does not include $c_5$, $c_7$, $c_{10}$;
the condition $v$) does not include $c_5$, $c_{10}$;
and the condition $vi$) does not include $c_5$, $c_7$.
Also, the condition $iv$) as well as $i$), $ii$), $vi$) require different conditions on $c_{10}$ which do not coincide with the DHOST condition in general.

\end{itemize}

\subsection{$c_1=c_2=0$ case}%%%%%%%%%%%%%%%%%%%%%
\label{app:case2}

In case 2, $c_1=c_2=0$ and the  kinetic term for $\zeta$ vanishes leaving $f_1=0$ and
$\delta N$ as the propagating degree of freedom.     
Since models where $\delta N$ and not $\zeta$ is propagating are unlikely to  recover Newtonian gravity,
we include this case for completeness and pedagogical interest.

\begin{itemize}
\item[a.]
Let us begin with the case $f_4f_7\ne 0$ and all other $f$'s zero.   In this case
we must again check for corrections to the dispersion relation beyond the
static limit.
Using only the condition $c_1=c_2=0$, the quadratic Lagrangian~\eqref{L2s-c} reduces to
\begin{align} 
\L_2 &= \f{1}{2}c_3 \dot{\delta N}^2 + \mk{c_4 + c_5 \f{k^2}{a^2}} \dot\zeta \delta N \notag\\
&~~~ + \f{1}{2}\mk{c_6 \f{k^2}{a^2} + c_7 \f{k^4}{a^4}} \zeta^2 + c_8 \f{k^2}{a^2} \zeta \delta N + \f{1}{2}\mk{c_9 + c_{10} \f{k^2}{a^2}} \delta N^2 \notag\\
&~~~ + \f{1}{2}c_{11} \f{k^4}{a^4} \psi^2 + \psi \mk{ \f{c_4}{3} \f{k^2}{a^2} \delta N + c_{12} \f{k^4}{a^4} \zeta } .
\end{align}
Here, again, the problematic term is $c_5 (k/a)^2 \dot\zeta \delta N$, and hence we impose $c_5=0$ as a supplemental degeneracy condition.

Requiring $f_2=f_3 =f_5=f_6=0$  leads to four possible cases
\begin{align} 
&i)&
  &c_1=c_2=0, &
  &c_5=c_{10}=0, &
  &c_3 = 0, & 
  &9c_9 (c_{12}^2 -c_7c_{11} ) + 9c_8^2 c_{11} + c_4^2 c_7 - 6 c_4 c_8 c_{12}=0,
  \notag\\
&ii)&
  &c_1=c_2=0, &
  &c_5=c_{10}=c_{12}^2-c_7 c_{11}= 0, &
  &c_6 = 0, & 
  &9c_8^2 c_{11} + c_4^2 c_7 - 6 c_4 c_8 c_{12}=0 , 
  \notag\\
&iii)&
  &c_1=c_2=0, &
  &c_5 =c_{12}^2-c_7c_{11}= 0, &
  &c_3=0, & 
  &9c_6 c_{10} c_{11} - 9c_8^2 c_{11} - c_4^2 c_7 + 6 c_4 c_8 c_{12}=0 , \notag\\
&iv)&
  &c_1=c_2=0, &
  &c_5 =c_{12}^2-c_7c_{11}= 0, &
  &c_6=0, & 
  &9c_8^2 c_{11} + c_4^2 c_7 - 6 c_4 c_8 c_{12}=0 .
\end{align}

\item[b.]
Next we consider the case $f_3f_6\ne 0$ and all other $f$'s zero.
Requiring $f_4=f_7=0$ under $f_3=c_{11} (c_3 c_6 - 2 c_4 c_5)\ne 0$ allows $c_4=c_9=0$ only.
By further requiring $f_2=f_5=0$, we obtain three possible cases
\begin{align} 
&i)&
  &c_1=c_2=0, &
  &c_4=c_9=0, &
  &c_{10}=c_3 (c_7 c_{11} - c_{12}^2) - c_5^2 c_{11} = 0  , &\notag\\
&ii)&
  &c_1=c_2=0, &
  &c_4=c_9=0, &
  &c_5 =c_{12}^2-c_7c_{11}= 0  , &\notag\\
&iii)&
  &c_1=c_2=0, &
  &c_4=c_9=0, &
  &c_{11}=c_{12}=0  . &
\end{align}

\item[c.] Finally, we consider $f_2f_5\ne 0$ and all other $f$'s zero.
This leads to five possible cases
\begin{align} 
&i)&
  &c_1=c_2=0, &
  &c_4=c_9=0, &
  &c_3=0, \hspace{10mm}
  c_6 c_{10} - c_8^2 = 0 , &\notag\\
&ii)&
  &c_1=c_2=0, &
  &c_4=c_9=0, &
  &c_6=0, \hspace{10mm}
  c_{11}(c_6 c_{10} - c_8^2) = 0 , &\notag\\
&iii)&
  &c_1=c_2=0, &
  &c_4=c_9=0, &
  &c_{11}=0, &\notag\\
&iv)&
  &c_1=c_2=0, &
  &c_4=c_6=0, &
  &c_9 (c_7c_{11} -c_{12}^2 )- c_8^2 c_{11} = 0 , &\notag\\
&v)&
  &c_1=c_2=0, &
  &c_6=c_{11}=0, &
  &9 c_9 c_{12}^2 + c_4^2 c_7 - 6 c_4 c_8 c_{12} = 0 . &
\end{align}

\end{itemize}

\subsection{$c_{11}=c_1/9$ case}%%%%%%%%%%%%%%%%%%%%%
\label{app:case3}

Finally we consider the case 3 where $c_{11}=c_1/9$.  Here the Euler-Lagrange equation for $\psi$ provides a contribution that cancels the  kinetic term for $\zeta$, and $\delta N$ is
again the propagating degree of freedom.      As in case 2, this case is unlikely to provide viable theories of gravity.
While generally we expect 5 static degeneracy
conditions, in this case there are fewer since some of the $f$ terms are identically zero
once other degeneracy conditions are applied.

\begin{itemize}

\item[a.]  The $f_4 f_7 \ne 0$ branch does not exist since $c_{11}=c_1/9$ implies $f_4=0$.

\item[b.]
Next, for $f_3f_6\ne 0$, 
requiring that additionally $f_2=f_5=f_7=0$ leads to two possible cases
\begin{align} 
&i)&
  &c_{11}=\f{c_1}{9}, &
  &c_4^2=c_1c_9, &
  &c_7 = \f{c_1 c_5^2 + 6 c_2 c_5 c_{12} + 9 c_3 c_{12}^2 }{c_1 c_3 - c_2^2}, &
  &c_{10}=0 , &\notag\\
&ii)&
  &c_{11}=\f{c_1}{9}, &
  &c_4^2=c_1c_9, &
  &c_7 = \f{c_1 c_5^2 + 6 c_2 c_5 c_{12} + 9 c_3 c_{12}^2 }{c_1 c_3 - c_2^2}, &
  &c_1 c_5 + 3c_2c_{12}=0 . &
\end{align}

\item[c.]
Finally $f_2f_5\ne 0$ leads to two possible cases
\begin{align} 
&i)&
  &c_{11}=\f{c_1}{9}, &
  &c_4^2=c_1c_9, &
  &c_2^2=c_1c_3 , \hspace{10mm}
  c_1 (c_6 c_{10} - c_8^2) + 6 c_4 c_8 c_{12} - 9 c_9c_{12}^2 = 0, &\notag\\
&ii)&
  &c_{11}=\f{c_1}{9}, &  
  &c_6=0, &  
  &c_9 = - \f{6 c_4 c_8 c_{12} - c_1 c_8^2 - c_4^2 c_7}{ c_1 c_7-9 c_{12}^2} .
\end{align}

\end{itemize}

\section{Relationship to literature}%%%%%%%%%%%%%%%%%%%%%%%%%%%%%%%%%%%%%%%%%
\label{app:rel}

Our approach is most similar to Ref.~\cite{Langlois:2017mxy} and in this Appendix we make the explicit
connection to that work and discuss the differences.    First some of the terms in Ref.~\cite{Langlois:2017mxy} take
a superficially different form that is related to ours through integration by parts. 
Up to a total derivative
\be \label{dKdRint} N\sqrt{h} \tilde\C_{KR} \delta K^i_j \delta R^j_i \sim \f{a^3}{2} \kk{ (\dot{\tilde\C}_{KR} + H\tilde\C_{KR})\mk{\f{\delta \sqrt{h}}{a^3} \delta R + \delta_2 R } + \tilde\C_{KR} \delta R\delta K + H\tilde\C_{KR} \delta N \delta R } , \ee 
and hence we can rewrite our quadratic Lagrangian \eqref{L2} in the form of Eq.~(1.2) of \cite{Langlois:2017mxy} 
expose the difference between the two
\begin{align} \label{L2comp} 
\delta \L_2  &\sim
 - a^3 \C_\beta\kk{ 
\delta N \mk{ \f{\delta_1\sqrt{h}}{a^3} }^\centerdot   
+ \delta N^i \delta N_{,i} } \notag\\
&~~~ + a^3\kk{ \C_{\beta R} \delta_1 R \dot{\delta N}
+ \f{1}{2} \C_{RR} \delta_1 R^2 
+ \f{1}{2} \tilde \C_{RR} \delta_1 R^i_j \delta_1 R^j_i 
+ \mk{ \C_{KR} + \f{1}{2} \tilde \C_{KR} } \delta K \delta_1 R 
} . 
\end{align}
The quadratic Lagrangian \eqref{L2} thus contains terms that differ from Eq.~(1.2) of \cite{Langlois:2017mxy}.  
Since the first term  in \eqref{L2comp} has $\delta N$ or $\delta N_{,i}$, 
it is nonvanishing only for scalar perturbations.
For scalar perturbation, it can be expressed up to a total derivative as
\begin{align} 
- a^3 \C_\beta \kk{ 
\delta N \mk{ \f{\delta_1\sqrt{h}}{a^3} }^\centerdot   
+ \delta N^i \delta N_{,i} } 
&\sim - a^3 \C_\beta \mk{ 3\dot\zeta - \f{\pa^2\psi}{a^2}} \delta N \notag\\
&= - a^3 \C_\beta ( \delta K + 3H\delta N ) \delta N , 
\end{align}
which can be absorbed into the $\alpha_B$ and $\alpha_K$ terms in Eq.~(1.2) of \cite{Langlois:2017mxy}. 
On the other hand, the third line of \eqref{L2comp} is not considered in \cite{Langlois:2017mxy} 
as these terms have derivatives higher than second order in total.
If we assume these terms are vanishing by imposing
\be \label{Lang-assump} \C_{\beta R} = \C_{RR} = \tilde \C_{RR} = \C_{KR} + \f{1}{2} \tilde \C_{KR} = 0, \ee
we have $c_5=c_7=c_{12}=0$ in \eqref{coeffs-2}.   These conditions hold in the 1a degeneracy subclasses
defined  by Eq.~(\ref{D1}) and (\ref{D2}).
Ref.~\cite{Langlois:2017mxy} considered only these cases. 
They furthermore assume $\phi \propto t$ and so their
\begin{equation}
V = -\frac{\dot\phi}{N} \beta
\end{equation}
vanishes in the background $\bar V=0$ or $\bar \beta=0$ in our notation.  Generalizing this does not change the functional 
form of their Lagrangian, just the mapping between the scalar field and ADM representations and so we retain $\bar\beta\ne 0$ in the correspondences below.  Note that if a field redefinition $\varphi \propto t(\phi)$ is performed instead after solving for the background $\phi(t)$,
which  alternately reestablishes the generality of their expressions, then the
DHOST coefficients must correspondingly be redefined (cf.~\cite{Crisostomi:2018bsp} v2).

In summary, in the subclass of \eqref{Lang-assump}, 
the quadratic Lagrangian~\eqref{L2} for scalar perturbation 
takes the same functional form as 
Eq.~(1.2) of \cite{Langlois:2017mxy} with the correspondence  
\begin{align} \label{dic4}
M^2 &= \tilde \C_{KK} ,\notag\\
\alpha_K &= \f{1}{H^2\tilde\C_{KK}} \mk{ 2\C_N + \C_{NN}  {- 2 \bar\beta \C_{\beta N} +\bar\beta^2\C_{\beta\beta} } + \f{(a^3\C_\beta)^\centerdot}{a^3} - \f{(a^3\C_{\beta N})^\centerdot}{a^3}  { + \f{(a^3\bar\beta\C_{\beta\beta})^\centerdot}{a^3}  } - 6 H \C_\beta },\notag\\
\alpha_B &= \f{\C_{NK}  { - \bar\beta\C_{\beta K} } - \C_\beta }{2H\tilde\C_{KK}} ,\notag\\
\alpha_T &= \f{2 \C_R + \dot {\tilde \C} _{KR} + H  \tilde \C_{KR}}{\tilde \C_{KK}} - 1 ,\notag\\
\alpha_H &= \f{2 ( \C_{NR} + \C_R) + H  \tilde \C_{KR}}{ \tilde \C_{KK}} - 1  ,\notag\\
\alpha_L &= - \f{3}{2} \mk{\f{\C_{KK}}{\tilde\C_{KK}} + 1} ,\notag\\
\beta_1 &= \f{\C_{\beta K}}{2\tilde \C_{KK}},\notag\\
\beta_2 &= \f{\C_{\beta\beta}}{\tilde \C_{KK}},\notag\\
\beta_3 &= \f{2\C_\alpha}{\tilde \C_{KK}}.
\end{align}
Equivalently, the inverse correspondence between notations for the subclass \eqref{Lang-assump} is given by
\begin{equation}
a^{-3} c_1 = -6M^2 (1+\alpha_L), \quad a^{-3} c_2 = 6  M^2\beta_1, \quad a^{-3}  c_3 = M^2 \beta_2 , \quad a^{-3} c_{10} = M^2 \beta_3 , \quad
a^{-3} c_{11} = -\frac{2}{3} M^2 \alpha_L ,
\end{equation}
and
\begin{align} 
\Theta &\equiv -\frac{1}{3} a^{-3} c_4 =  -2 H M^2(1+ \alpha_B + \alpha_L) , \nonumber\\
\Psi &\equiv \frac{1}{4} a^{-3} c_6 =  \frac{1}{2} M^2 (1+\alpha_T) , \nonumber\\
\Xi &\equiv  \frac{1}{4} a^{-3} c_8 = \frac{1}{2} M^2 (1+\alpha_H) , \nonumber\\
\Phi &\equiv a^{-3} c_9 = H^2 M^2 [\alpha_K - 6(1+\alpha_L) - 12 \alpha_B ]+ 6 a^{-3} \left( a^3 M^2 H \beta_1\right)^\centerdot . 
\end{align}
with $c_5,c_7,c_{12}$ vanishing in this class.

With these relations we can also translate the degeneracy conditions Eqs.~(2.15), (2.16) of \cite{Langlois:2017mxy}:
\begin{align}
{\rm C}_{\rm I}:& \quad 
\alpha_L = 0, \quad 
\beta_2 = -6 \beta_1^2, \quad 
\beta_3 = -2 \beta_1 [2 (1 + \alpha_H) + \beta_1 (1 + \alpha_T)] ,\\
{\rm C}_{\rm II}:& \quad 
\beta_1 = -(1 + \alpha_L) \f{1 + \alpha_H}{1 + \alpha_T}, \quad
\beta_2 = -6 (1 + \alpha_L) \f{(1 + \alpha_H)^2}{(1 + \alpha_T)^2}, \quad
\beta_3 = 2 \f{(1 + \alpha_H)^2}{1 + \alpha_T} ,
\end{align}
into our notation to confirm that  their ${\rm C}_{\rm I}$ and ${\rm C}_{\rm II}$ correspond to \eqref{D1} and \eqref{D2}, respectively.  The
Lagrangian for $\tilde\zeta$ in \eqref{S2s} is equivalent to Eq.~(4.8) of \cite{Langlois:2017mxy} for ${\rm C}_{\rm I}$.

\bibliography{ref-DegEFT}

\begin{thebibliography}{49}%
\makeatletter
\providecommand \@ifxundefined [1]{%
 \@ifx{#1\undefined}
}%
\providecommand \@ifnum [1]{%
 \ifnum #1\expandafter \@firstoftwo
 \else \expandafter \@secondoftwo
 \fi
}%
\providecommand \@ifx [1]{%
 \ifx #1\expandafter \@firstoftwo
 \else \expandafter \@secondoftwo
 \fi
}%
\providecommand \natexlab [1]{#1}%
\providecommand \enquote  [1]{``#1''}%
\providecommand \bibnamefont  [1]{#1}%
\providecommand \bibfnamefont [1]{#1}%
\providecommand \citenamefont [1]{#1}%
\providecommand \href@noop [0]{\@secondoftwo}%
\providecommand \href [0]{\begingroup \@sanitize@url \@href}%
\providecommand \@href[1]{\@@startlink{#1}\@@href}%
\providecommand \@@href[1]{\endgroup#1\@@endlink}%
\providecommand \@sanitize@url [0]{\catcode `\\12\catcode `\$12\catcode
  `\&12\catcode `\#12\catcode `\^12\catcode `\_12\catcode `\%12\relax}%
\providecommand \@@startlink[1]{}%
\providecommand \@@endlink[0]{}%
\providecommand \url  [0]{\begingroup\@sanitize@url \@url }%
\providecommand \@url [1]{\endgroup\@href {#1}{\urlprefix }}%
\providecommand \urlprefix  [0]{URL }%
\providecommand \Eprint [0]{\href }%
\providecommand \doibase [0]{http://dx.doi.org/}%
\providecommand \selectlanguage [0]{\@gobble}%
\providecommand \bibinfo  [0]{\@secondoftwo}%
\providecommand \bibfield  [0]{\@secondoftwo}%
\providecommand \translation [1]{[#1]}%
\providecommand \BibitemOpen [0]{}%
\providecommand \bibitemStop [0]{}%
\providecommand \bibitemNoStop [0]{.\EOS\space}%
\providecommand \EOS [0]{\spacefactor3000\relax}%
\providecommand \BibitemShut  [1]{\csname bibitem#1\endcsname}%
\let\auto@bib@innerbib\@empty
%</preamble>
\bibitem [{\citenamefont {Creminelli}\ \emph {et~al.}(2006)\citenamefont
  {Creminelli}, \citenamefont {Luty}, \citenamefont {Nicolis},\ and\
  \citenamefont {Senatore}}]{Creminelli:2006xe}%
  \BibitemOpen
  \bibfield  {author} {\bibinfo {author} {\bibfnamefont {P.}~\bibnamefont
  {Creminelli}}, \bibinfo {author} {\bibfnamefont {M.~A.}\ \bibnamefont
  {Luty}}, \bibinfo {author} {\bibfnamefont {A.}~\bibnamefont {Nicolis}}, \
  and\ \bibinfo {author} {\bibfnamefont {L.}~\bibnamefont {Senatore}},\ }\href
  {\doibase 10.1088/1126-6708/2006/12/080} {\bibfield  {journal} {\bibinfo
  {journal} {JHEP}\ }\textbf {\bibinfo {volume} {12}},\ \bibinfo {pages} {080}
  (\bibinfo {year} {2006})},\ \Eprint {http://arxiv.org/abs/hep-th/0606090}
  {arXiv:hep-th/0606090 [hep-th]} \BibitemShut {NoStop}%
%%CITATION = HEP-TH/0606090;%%
\bibitem [{\citenamefont {Cheung}\ \emph {et~al.}(2008)\citenamefont {Cheung},
  \citenamefont {Creminelli}, \citenamefont {Fitzpatrick}, \citenamefont
  {Kaplan},\ and\ \citenamefont {Senatore}}]{Cheung:2007st}%
  \BibitemOpen
  \bibfield  {author} {\bibinfo {author} {\bibfnamefont {C.}~\bibnamefont
  {Cheung}}, \bibinfo {author} {\bibfnamefont {P.}~\bibnamefont {Creminelli}},
  \bibinfo {author} {\bibfnamefont {A.~L.}\ \bibnamefont {Fitzpatrick}},
  \bibinfo {author} {\bibfnamefont {J.}~\bibnamefont {Kaplan}}, \ and\ \bibinfo
  {author} {\bibfnamefont {L.}~\bibnamefont {Senatore}},\ }\href {\doibase
  10.1088/1126-6708/2008/03/014} {\bibfield  {journal} {\bibinfo  {journal}
  {JHEP}\ }\textbf {\bibinfo {volume} {03}},\ \bibinfo {pages} {014} (\bibinfo
  {year} {2008})},\ \Eprint {http://arxiv.org/abs/0709.0293} {arXiv:0709.0293
  [hep-th]} \BibitemShut {NoStop}%
%%CITATION = ARXIV:0709.0293;%%
\bibitem [{\citenamefont {Gleyzes}\ \emph {et~al.}(2013)\citenamefont
  {Gleyzes}, \citenamefont {Langlois}, \citenamefont {Piazza},\ and\
  \citenamefont {Vernizzi}}]{Gleyzes:2013ooa}%
  \BibitemOpen
  \bibfield  {author} {\bibinfo {author} {\bibfnamefont {J.}~\bibnamefont
  {Gleyzes}}, \bibinfo {author} {\bibfnamefont {D.}~\bibnamefont {Langlois}},
  \bibinfo {author} {\bibfnamefont {F.}~\bibnamefont {Piazza}}, \ and\ \bibinfo
  {author} {\bibfnamefont {F.}~\bibnamefont {Vernizzi}},\ }\href {\doibase
  10.1088/1475-7516/2013/08/025} {\bibfield  {journal} {\bibinfo  {journal}
  {JCAP}\ }\textbf {\bibinfo {volume} {1308}},\ \bibinfo {pages} {025}
  (\bibinfo {year} {2013})},\ \Eprint {http://arxiv.org/abs/1304.4840}
  {arXiv:1304.4840 [hep-th]} \BibitemShut {NoStop}%
%%CITATION = ARXIV:1304.4840;%%
\bibitem [{\citenamefont {Kase}\ and\ \citenamefont
  {Tsujikawa}(2014)}]{Kase:2014cwa}%
  \BibitemOpen
  \bibfield  {author} {\bibinfo {author} {\bibfnamefont {R.}~\bibnamefont
  {Kase}}\ and\ \bibinfo {author} {\bibfnamefont {S.}~\bibnamefont
  {Tsujikawa}},\ }\href {\doibase 10.1142/S0218271814430081} {\bibfield
  {journal} {\bibinfo  {journal} {Int. J. Mod. Phys.}\ }\textbf {\bibinfo
  {volume} {D23}},\ \bibinfo {pages} {1443008} (\bibinfo {year} {2014})},\
  \Eprint {http://arxiv.org/abs/1409.1984} {arXiv:1409.1984 [hep-th]}
  \BibitemShut {NoStop}%
%%CITATION = ARXIV:1409.1984;%%
\bibitem [{\citenamefont {Gleyzes}\ \emph {et~al.}(2014)\citenamefont
  {Gleyzes}, \citenamefont {Langlois},\ and\ \citenamefont
  {Vernizzi}}]{Gleyzes:2014rba}%
  \BibitemOpen
  \bibfield  {author} {\bibinfo {author} {\bibfnamefont {J.}~\bibnamefont
  {Gleyzes}}, \bibinfo {author} {\bibfnamefont {D.}~\bibnamefont {Langlois}}, \
  and\ \bibinfo {author} {\bibfnamefont {F.}~\bibnamefont {Vernizzi}},\ }\href
  {\doibase 10.1142/S021827181443010X} {\bibfield  {journal} {\bibinfo
  {journal} {Int. J. Mod. Phys.}\ }\textbf {\bibinfo {volume} {D23}},\ \bibinfo
  {pages} {1443010} (\bibinfo {year} {2014})},\ \Eprint
  {http://arxiv.org/abs/1411.3712} {arXiv:1411.3712 [hep-th]} \BibitemShut
  {NoStop}%
%%CITATION = ARXIV:1411.3712;%%
\bibitem [{\citenamefont {Gleyzes}\ \emph
  {et~al.}(2015{\natexlab{a}})\citenamefont {Gleyzes}, \citenamefont
  {Langlois}, \citenamefont {Mancarella},\ and\ \citenamefont
  {Vernizzi}}]{Gleyzes:2015pma}%
  \BibitemOpen
  \bibfield  {author} {\bibinfo {author} {\bibfnamefont {J.}~\bibnamefont
  {Gleyzes}}, \bibinfo {author} {\bibfnamefont {D.}~\bibnamefont {Langlois}},
  \bibinfo {author} {\bibfnamefont {M.}~\bibnamefont {Mancarella}}, \ and\
  \bibinfo {author} {\bibfnamefont {F.}~\bibnamefont {Vernizzi}},\ }\href
  {\doibase 10.1088/1475-7516/2015/08/054} {\bibfield  {journal} {\bibinfo
  {journal} {JCAP}\ }\textbf {\bibinfo {volume} {1508}},\ \bibinfo {pages}
  {054} (\bibinfo {year} {2015}{\natexlab{a}})},\ \Eprint
  {http://arxiv.org/abs/1504.05481} {arXiv:1504.05481 [astro-ph.CO]}
  \BibitemShut {NoStop}%
%%CITATION = ARXIV:1504.05481;%%
\bibitem [{\citenamefont {Motohashi}\ and\ \citenamefont
  {Hu}(2017)}]{Motohashi:2017gqb}%
  \BibitemOpen
  \bibfield  {author} {\bibinfo {author} {\bibfnamefont {H.}~\bibnamefont
  {Motohashi}}\ and\ \bibinfo {author} {\bibfnamefont {W.}~\bibnamefont {Hu}},\
  }\href {\doibase 10.1103/PhysRevD.96.023502} {\bibfield  {journal} {\bibinfo
  {journal} {Phys. Rev.}\ }\textbf {\bibinfo {volume} {D96}},\ \bibinfo {pages}
  {023502} (\bibinfo {year} {2017})},\ \Eprint
  {http://arxiv.org/abs/1704.01128} {arXiv:1704.01128 [hep-th]} \BibitemShut
  {NoStop}%
%%CITATION = ARXIV:1704.01128;%%
\bibitem [{\citenamefont {Horndeski}(1974)}]{Horndeski:1974wa}%
  \BibitemOpen
  \bibfield  {author} {\bibinfo {author} {\bibfnamefont {G.~W.}\ \bibnamefont
  {Horndeski}},\ }\href {\doibase 10.1007/BF01807638} {\bibfield  {journal}
  {\bibinfo  {journal} {Int. J. Theor. Phys.}\ }\textbf {\bibinfo {volume}
  {10}},\ \bibinfo {pages} {363} (\bibinfo {year} {1974})}\BibitemShut
  {NoStop}%
%%CITATION = IJTPB,10,363;%%
\bibitem [{\citenamefont {Nicolis}\ \emph {et~al.}(2009)\citenamefont
  {Nicolis}, \citenamefont {Rattazzi},\ and\ \citenamefont
  {Trincherini}}]{Nicolis:2008in}%
  \BibitemOpen
  \bibfield  {author} {\bibinfo {author} {\bibfnamefont {A.}~\bibnamefont
  {Nicolis}}, \bibinfo {author} {\bibfnamefont {R.}~\bibnamefont {Rattazzi}}, \
  and\ \bibinfo {author} {\bibfnamefont {E.}~\bibnamefont {Trincherini}},\
  }\href {\doibase 10.1103/PhysRevD.79.064036} {\bibfield  {journal} {\bibinfo
  {journal} {Phys. Rev.}\ }\textbf {\bibinfo {volume} {D79}},\ \bibinfo {pages}
  {064036} (\bibinfo {year} {2009})},\ \Eprint {http://arxiv.org/abs/0811.2197}
  {arXiv:0811.2197 [hep-th]} \BibitemShut {NoStop}%
%%CITATION = ARXIV:0811.2197;%%
\bibitem [{\citenamefont {Deffayet}\ \emph
  {et~al.}(2009{\natexlab{a}})\citenamefont {Deffayet}, \citenamefont
  {Esposito-Farese},\ and\ \citenamefont {Vikman}}]{Deffayet:2009wt}%
  \BibitemOpen
  \bibfield  {author} {\bibinfo {author} {\bibfnamefont {C.}~\bibnamefont
  {Deffayet}}, \bibinfo {author} {\bibfnamefont {G.}~\bibnamefont
  {Esposito-Farese}}, \ and\ \bibinfo {author} {\bibfnamefont {A.}~\bibnamefont
  {Vikman}},\ }\href {\doibase 10.1103/PhysRevD.79.084003} {\bibfield
  {journal} {\bibinfo  {journal} {Phys. Rev.}\ }\textbf {\bibinfo {volume}
  {D79}},\ \bibinfo {pages} {084003} (\bibinfo {year} {2009}{\natexlab{a}})},\
  \Eprint {http://arxiv.org/abs/0901.1314} {arXiv:0901.1314 [hep-th]}
  \BibitemShut {NoStop}%
%%CITATION = ARXIV:0901.1314;%%
\bibitem [{\citenamefont {Deffayet}\ \emph
  {et~al.}(2009{\natexlab{b}})\citenamefont {Deffayet}, \citenamefont {Deser},\
  and\ \citenamefont {Esposito-Farese}}]{Deffayet:2009mn}%
  \BibitemOpen
  \bibfield  {author} {\bibinfo {author} {\bibfnamefont {C.}~\bibnamefont
  {Deffayet}}, \bibinfo {author} {\bibfnamefont {S.}~\bibnamefont {Deser}}, \
  and\ \bibinfo {author} {\bibfnamefont {G.}~\bibnamefont {Esposito-Farese}},\
  }\href {\doibase 10.1103/PhysRevD.80.064015} {\bibfield  {journal} {\bibinfo
  {journal} {Phys. Rev.}\ }\textbf {\bibinfo {volume} {D80}},\ \bibinfo {pages}
  {064015} (\bibinfo {year} {2009}{\natexlab{b}})},\ \Eprint
  {http://arxiv.org/abs/0906.1967} {arXiv:0906.1967 [gr-qc]} \BibitemShut
  {NoStop}%
%%CITATION = ARXIV:0906.1967;%%
\bibitem [{\citenamefont {Deffayet}\ \emph {et~al.}(2011)\citenamefont
  {Deffayet}, \citenamefont {Gao}, \citenamefont {Steer},\ and\ \citenamefont
  {Zahariade}}]{Deffayet:2011gz}%
  \BibitemOpen
  \bibfield  {author} {\bibinfo {author} {\bibfnamefont {C.}~\bibnamefont
  {Deffayet}}, \bibinfo {author} {\bibfnamefont {X.}~\bibnamefont {Gao}},
  \bibinfo {author} {\bibfnamefont {D.~A.}\ \bibnamefont {Steer}}, \ and\
  \bibinfo {author} {\bibfnamefont {G.}~\bibnamefont {Zahariade}},\ }\href
  {\doibase 10.1103/PhysRevD.84.064039} {\bibfield  {journal} {\bibinfo
  {journal} {Phys. Rev.}\ }\textbf {\bibinfo {volume} {D84}},\ \bibinfo {pages}
  {064039} (\bibinfo {year} {2011})},\ \Eprint {http://arxiv.org/abs/1103.3260}
  {arXiv:1103.3260 [hep-th]} \BibitemShut {NoStop}%
%%CITATION = ARXIV:1103.3260;%%
\bibitem [{\citenamefont {Kobayashi}\ \emph {et~al.}(2011)\citenamefont
  {Kobayashi}, \citenamefont {Yamaguchi},\ and\ \citenamefont
  {Yokoyama}}]{Kobayashi:2011nu}%
  \BibitemOpen
  \bibfield  {author} {\bibinfo {author} {\bibfnamefont {T.}~\bibnamefont
  {Kobayashi}}, \bibinfo {author} {\bibfnamefont {M.}~\bibnamefont
  {Yamaguchi}}, \ and\ \bibinfo {author} {\bibfnamefont {J.}~\bibnamefont
  {Yokoyama}},\ }\href {\doibase 10.1143/PTP.126.511} {\bibfield  {journal}
  {\bibinfo  {journal} {Prog. Theor. Phys.}\ }\textbf {\bibinfo {volume}
  {126}},\ \bibinfo {pages} {511} (\bibinfo {year} {2011})},\ \Eprint
  {http://arxiv.org/abs/1105.5723} {arXiv:1105.5723 [hep-th]} \BibitemShut
  {NoStop}%
%%CITATION = ARXIV:1105.5723;%%
\bibitem [{\citenamefont {Gleyzes}\ \emph
  {et~al.}(2015{\natexlab{b}})\citenamefont {Gleyzes}, \citenamefont
  {Langlois}, \citenamefont {Piazza},\ and\ \citenamefont
  {Vernizzi}}]{Gleyzes:2014dya}%
  \BibitemOpen
  \bibfield  {author} {\bibinfo {author} {\bibfnamefont {J.}~\bibnamefont
  {Gleyzes}}, \bibinfo {author} {\bibfnamefont {D.}~\bibnamefont {Langlois}},
  \bibinfo {author} {\bibfnamefont {F.}~\bibnamefont {Piazza}}, \ and\ \bibinfo
  {author} {\bibfnamefont {F.}~\bibnamefont {Vernizzi}},\ }\href {\doibase
  10.1103/PhysRevLett.114.211101} {\bibfield  {journal} {\bibinfo  {journal}
  {Phys. Rev. Lett.}\ }\textbf {\bibinfo {volume} {114}},\ \bibinfo {pages}
  {211101} (\bibinfo {year} {2015}{\natexlab{b}})},\ \Eprint
  {http://arxiv.org/abs/1404.6495} {arXiv:1404.6495 [hep-th]} \BibitemShut
  {NoStop}%
%%CITATION = ARXIV:1404.6495;%%
\bibitem [{\citenamefont {Gleyzes}\ \emph
  {et~al.}(2015{\natexlab{c}})\citenamefont {Gleyzes}, \citenamefont
  {Langlois}, \citenamefont {Piazza},\ and\ \citenamefont
  {Vernizzi}}]{Gleyzes:2014qga}%
  \BibitemOpen
  \bibfield  {author} {\bibinfo {author} {\bibfnamefont {J.}~\bibnamefont
  {Gleyzes}}, \bibinfo {author} {\bibfnamefont {D.}~\bibnamefont {Langlois}},
  \bibinfo {author} {\bibfnamefont {F.}~\bibnamefont {Piazza}}, \ and\ \bibinfo
  {author} {\bibfnamefont {F.}~\bibnamefont {Vernizzi}},\ }\href {\doibase
  10.1088/1475-7516/2015/02/018} {\bibfield  {journal} {\bibinfo  {journal}
  {JCAP}\ }\textbf {\bibinfo {volume} {1502}},\ \bibinfo {pages} {018}
  (\bibinfo {year} {2015}{\natexlab{c}})},\ \Eprint
  {http://arxiv.org/abs/1408.1952} {arXiv:1408.1952 [astro-ph.CO]} \BibitemShut
  {NoStop}%
%%CITATION = ARXIV:1408.1952;%%
\bibitem [{\citenamefont {Horava}(2009)}]{Horava:2009uw}%
  \BibitemOpen
  \bibfield  {author} {\bibinfo {author} {\bibfnamefont {P.}~\bibnamefont
  {Horava}},\ }\href {\doibase 10.1103/PhysRevD.79.084008} {\bibfield
  {journal} {\bibinfo  {journal} {Phys. Rev.}\ }\textbf {\bibinfo {volume}
  {D79}},\ \bibinfo {pages} {084008} (\bibinfo {year} {2009})},\ \Eprint
  {http://arxiv.org/abs/0901.3775} {arXiv:0901.3775 [hep-th]} \BibitemShut
  {NoStop}%
%%CITATION = ARXIV:0901.3775;%%
\bibitem [{\citenamefont {Blas}\ \emph
  {et~al.}(2010{\natexlab{a}})\citenamefont {Blas}, \citenamefont {Pujolas},\
  and\ \citenamefont {Sibiryakov}}]{Blas:2009qj}%
  \BibitemOpen
  \bibfield  {author} {\bibinfo {author} {\bibfnamefont {D.}~\bibnamefont
  {Blas}}, \bibinfo {author} {\bibfnamefont {O.}~\bibnamefont {Pujolas}}, \
  and\ \bibinfo {author} {\bibfnamefont {S.}~\bibnamefont {Sibiryakov}},\
  }\href {\doibase 10.1103/PhysRevLett.104.181302} {\bibfield  {journal}
  {\bibinfo  {journal} {Phys. Rev. Lett.}\ }\textbf {\bibinfo {volume} {104}},\
  \bibinfo {pages} {181302} (\bibinfo {year} {2010}{\natexlab{a}})},\ \Eprint
  {http://arxiv.org/abs/0909.3525} {arXiv:0909.3525 [hep-th]} \BibitemShut
  {NoStop}%
%%CITATION = ARXIV:0909.3525;%%
\bibitem [{\citenamefont {Blas}\ \emph
  {et~al.}(2010{\natexlab{b}})\citenamefont {Blas}, \citenamefont {Pujolas},\
  and\ \citenamefont {Sibiryakov}}]{Blas:2009ck}%
  \BibitemOpen
  \bibfield  {author} {\bibinfo {author} {\bibfnamefont {D.}~\bibnamefont
  {Blas}}, \bibinfo {author} {\bibfnamefont {O.}~\bibnamefont {Pujolas}}, \
  and\ \bibinfo {author} {\bibfnamefont {S.}~\bibnamefont {Sibiryakov}},\
  }\href {\doibase 10.1016/j.physletb.2010.03.073} {\bibfield  {journal}
  {\bibinfo  {journal} {Phys. Lett.}\ }\textbf {\bibinfo {volume} {B688}},\
  \bibinfo {pages} {350} (\bibinfo {year} {2010}{\natexlab{b}})},\ \Eprint
  {http://arxiv.org/abs/0912.0550} {arXiv:0912.0550 [hep-th]} \BibitemShut
  {NoStop}%
%%CITATION = ARXIV:0912.0550;%%
\bibitem [{\citenamefont {Ostrogradsky}(1850)}]{Ostrogradsky:1850fid}%
  \BibitemOpen
  \bibfield  {author} {\bibinfo {author} {\bibfnamefont {M.}~\bibnamefont
  {Ostrogradsky}},\ }\href
  {http://hdl.handle.net/2027/mdp.39015038710128?urlappend=%3Bseq=405}
  {\bibfield  {journal} {\bibinfo  {journal} {Mem. Acad. St. Petersbourg}\
  }\textbf {\bibinfo {volume} {6}},\ \bibinfo {pages} {385} (\bibinfo {year}
  {1850})}\BibitemShut {NoStop}%
%%CITATION = INSPIRE-1468685;%%
\bibitem [{\citenamefont {Woodard}(2015)}]{Woodard:2015zca}%
  \BibitemOpen
  \bibfield  {author} {\bibinfo {author} {\bibfnamefont {R.~P.}\ \bibnamefont
  {Woodard}},\ }\href {\doibase 10.4249/scholarpedia.32243} {\bibfield
  {journal} {\bibinfo  {journal} {Scholarpedia}\ }\textbf {\bibinfo {volume}
  {10}},\ \bibinfo {pages} {32243} (\bibinfo {year} {2015})},\ \Eprint
  {http://arxiv.org/abs/1506.02210} {arXiv:1506.02210 [hep-th]} \BibitemShut
  {NoStop}%
%%CITATION = ARXIV:1506.02210;%%
\bibitem [{\citenamefont {Raidal}\ and\ \citenamefont
  {Veermäe}(2017)}]{Raidal:2016wop}%
  \BibitemOpen
  \bibfield  {author} {\bibinfo {author} {\bibfnamefont {M.}~\bibnamefont
  {Raidal}}\ and\ \bibinfo {author} {\bibfnamefont {H.}~\bibnamefont
  {Veermäe}},\ }\href {\doibase 10.1016/j.nuclphysb.2017.01.024} {\bibfield
  {journal} {\bibinfo  {journal} {Nucl.\ Phys.\ B}\ }\textbf {\bibinfo {volume}
  {916}},\ \bibinfo {pages} {607} (\bibinfo {year} {2017})},\ \Eprint
  {http://arxiv.org/abs/1611.03498} {arXiv:1611.03498 [hep-th]} \BibitemShut
  {NoStop}%
\bibitem [{\citenamefont {Smilga}(2017)}]{Smilga:2017arl}%
  \BibitemOpen
  \bibfield  {author} {\bibinfo {author} {\bibfnamefont {A.}~\bibnamefont
  {Smilga}},\ }\href {\doibase 10.1142/S0217751X17300253} {\bibfield  {journal}
  {\bibinfo  {journal} {Int.\ J.\ Mod.\ Phys.\ A}\ }\textbf {\bibinfo {volume}
  {32}},\ \bibinfo {pages} {1730025} (\bibinfo {year} {2017})},\ \Eprint
  {http://arxiv.org/abs/1710.11538} {arXiv:1710.11538 [hep-th]} \BibitemShut
  {NoStop}%
\bibitem [{\citenamefont {Motohashi}\ and\ \citenamefont
  {Suyama}(2020)}]{Motohashi:2020psc}%
  \BibitemOpen
  \bibfield  {author} {\bibinfo {author} {\bibfnamefont {H.}~\bibnamefont
  {Motohashi}}\ and\ \bibinfo {author} {\bibfnamefont {T.}~\bibnamefont
  {Suyama}},\ }\href@noop {} {\  (\bibinfo {year} {2020})},\ \Eprint
  {http://arxiv.org/abs/2001.02483} {arXiv:2001.02483 [hep-th]} \BibitemShut
  {NoStop}%
%%CITATION = ARXIV:2001.02483;%%
\bibitem [{\citenamefont {Motohashi}\ and\ \citenamefont
  {Suyama}(2015)}]{Motohashi:2014opa}%
  \BibitemOpen
  \bibfield  {author} {\bibinfo {author} {\bibfnamefont {H.}~\bibnamefont
  {Motohashi}}\ and\ \bibinfo {author} {\bibfnamefont {T.}~\bibnamefont
  {Suyama}},\ }\href {\doibase 10.1103/PhysRevD.91.085009} {\bibfield
  {journal} {\bibinfo  {journal} {Phys. Rev.}\ }\textbf {\bibinfo {volume}
  {D91}},\ \bibinfo {pages} {085009} (\bibinfo {year} {2015})},\ \Eprint
  {http://arxiv.org/abs/1411.3721} {arXiv:1411.3721 [physics.class-ph]}
  \BibitemShut {NoStop}%
%%CITATION = ARXIV:1411.3721;%%
\bibitem [{\citenamefont {Langlois}\ and\ \citenamefont
  {Noui}(2016)}]{Langlois:2015cwa}%
  \BibitemOpen
  \bibfield  {author} {\bibinfo {author} {\bibfnamefont {D.}~\bibnamefont
  {Langlois}}\ and\ \bibinfo {author} {\bibfnamefont {K.}~\bibnamefont
  {Noui}},\ }\href {\doibase 10.1088/1475-7516/2016/02/034} {\bibfield
  {journal} {\bibinfo  {journal} {JCAP}\ }\textbf {\bibinfo {volume} {1602}},\
  \bibinfo {pages} {034} (\bibinfo {year} {2016})},\ \Eprint
  {http://arxiv.org/abs/1510.06930} {arXiv:1510.06930 [gr-qc]} \BibitemShut
  {NoStop}%
%%CITATION = ARXIV:1510.06930;%%
\bibitem [{\citenamefont {Motohashi}\ \emph {et~al.}(2016)\citenamefont
  {Motohashi}, \citenamefont {Noui}, \citenamefont {Suyama}, \citenamefont
  {Yamaguchi},\ and\ \citenamefont {Langlois}}]{Motohashi:2016ftl}%
  \BibitemOpen
  \bibfield  {author} {\bibinfo {author} {\bibfnamefont {H.}~\bibnamefont
  {Motohashi}}, \bibinfo {author} {\bibfnamefont {K.}~\bibnamefont {Noui}},
  \bibinfo {author} {\bibfnamefont {T.}~\bibnamefont {Suyama}}, \bibinfo
  {author} {\bibfnamefont {M.}~\bibnamefont {Yamaguchi}}, \ and\ \bibinfo
  {author} {\bibfnamefont {D.}~\bibnamefont {Langlois}},\ }\href {\doibase
  10.1088/1475-7516/2016/07/033} {\bibfield  {journal} {\bibinfo  {journal}
  {JCAP}\ }\textbf {\bibinfo {volume} {1607}},\ \bibinfo {pages} {033}
  (\bibinfo {year} {2016})},\ \Eprint {http://arxiv.org/abs/1603.09355}
  {arXiv:1603.09355 [hep-th]} \BibitemShut {NoStop}%
%%CITATION = ARXIV:1603.09355;%%
\bibitem [{\citenamefont {Motohashi}\ \emph
  {et~al.}(2018{\natexlab{a}})\citenamefont {Motohashi}, \citenamefont
  {Suyama},\ and\ \citenamefont {Yamaguchi}}]{Motohashi:2017eya}%
  \BibitemOpen
  \bibfield  {author} {\bibinfo {author} {\bibfnamefont {H.}~\bibnamefont
  {Motohashi}}, \bibinfo {author} {\bibfnamefont {T.}~\bibnamefont {Suyama}}, \
  and\ \bibinfo {author} {\bibfnamefont {M.}~\bibnamefont {Yamaguchi}},\ }\href
  {\doibase 10.7566/JPSJ.87.063401} {\bibfield  {journal} {\bibinfo  {journal}
  {J. Phys. Soc. Jap.}\ }\textbf {\bibinfo {volume} {87}},\ \bibinfo {pages}
  {063401} (\bibinfo {year} {2018}{\natexlab{a}})},\ \Eprint
  {http://arxiv.org/abs/1711.08125} {arXiv:1711.08125 [hep-th]} \BibitemShut
  {NoStop}%
%%CITATION = ARXIV:1711.08125;%%
\bibitem [{\citenamefont {Motohashi}\ \emph
  {et~al.}(2018{\natexlab{b}})\citenamefont {Motohashi}, \citenamefont
  {Suyama},\ and\ \citenamefont {Yamaguchi}}]{Motohashi:2018pxg}%
  \BibitemOpen
  \bibfield  {author} {\bibinfo {author} {\bibfnamefont {H.}~\bibnamefont
  {Motohashi}}, \bibinfo {author} {\bibfnamefont {T.}~\bibnamefont {Suyama}}, \
  and\ \bibinfo {author} {\bibfnamefont {M.}~\bibnamefont {Yamaguchi}},\ }\href
  {\doibase 10.1007/JHEP06(2018)133} {\bibfield  {journal} {\bibinfo  {journal}
  {JHEP}\ }\textbf {\bibinfo {volume} {06}},\ \bibinfo {pages} {133} (\bibinfo
  {year} {2018}{\natexlab{b}})},\ \Eprint {http://arxiv.org/abs/1804.07990}
  {arXiv:1804.07990 [hep-th]} \BibitemShut {NoStop}%
%%CITATION = ARXIV:1804.07990;%%
\bibitem [{\citenamefont {Aoki}\ and\ \citenamefont
  {Motohashi}(2020)}]{Aoki:2020gfv}%
  \BibitemOpen
  \bibfield  {author} {\bibinfo {author} {\bibfnamefont {K.}~\bibnamefont
  {Aoki}}\ and\ \bibinfo {author} {\bibfnamefont {H.}~\bibnamefont
  {Motohashi}},\ }\href@noop {} {\  (\bibinfo {year} {2020})},\ \Eprint
  {http://arxiv.org/abs/2001.06756} {arXiv:2001.06756 [hep-th]} \BibitemShut
  {NoStop}%
%%CITATION = ARXIV:2001.06756;%%
\bibitem [{\citenamefont {Ben~Achour}\ \emph {et~al.}(2016)\citenamefont
  {Ben~Achour}, \citenamefont {Crisostomi}, \citenamefont {Koyama},
  \citenamefont {Langlois}, \citenamefont {Noui},\ and\ \citenamefont
  {Tasinato}}]{BenAchour:2016fzp}%
  \BibitemOpen
  \bibfield  {author} {\bibinfo {author} {\bibfnamefont {J.}~\bibnamefont
  {Ben~Achour}}, \bibinfo {author} {\bibfnamefont {M.}~\bibnamefont
  {Crisostomi}}, \bibinfo {author} {\bibfnamefont {K.}~\bibnamefont {Koyama}},
  \bibinfo {author} {\bibfnamefont {D.}~\bibnamefont {Langlois}}, \bibinfo
  {author} {\bibfnamefont {K.}~\bibnamefont {Noui}}, \ and\ \bibinfo {author}
  {\bibfnamefont {G.}~\bibnamefont {Tasinato}},\ }\href {\doibase
  10.1007/JHEP12(2016)100} {\bibfield  {journal} {\bibinfo  {journal} {JHEP}\
  }\textbf {\bibinfo {volume} {12}},\ \bibinfo {pages} {100} (\bibinfo {year}
  {2016})},\ \Eprint {http://arxiv.org/abs/1608.08135} {arXiv:1608.08135
  [hep-th]} \BibitemShut {NoStop}%
%%CITATION = ARXIV:1608.08135;%%
\bibitem [{\citenamefont {Langlois}\ \emph {et~al.}(2017)\citenamefont
  {Langlois}, \citenamefont {Mancarella}, \citenamefont {Noui},\ and\
  \citenamefont {Vernizzi}}]{Langlois:2017mxy}%
  \BibitemOpen
  \bibfield  {author} {\bibinfo {author} {\bibfnamefont {D.}~\bibnamefont
  {Langlois}}, \bibinfo {author} {\bibfnamefont {M.}~\bibnamefont
  {Mancarella}}, \bibinfo {author} {\bibfnamefont {K.}~\bibnamefont {Noui}}, \
  and\ \bibinfo {author} {\bibfnamefont {F.}~\bibnamefont {Vernizzi}},\ }\href
  {\doibase 10.1088/1475-7516/2017/05/033} {\bibfield  {journal} {\bibinfo
  {journal} {JCAP}\ }\textbf {\bibinfo {volume} {1705}},\ \bibinfo {pages}
  {033} (\bibinfo {year} {2017})},\ \Eprint {http://arxiv.org/abs/1703.03797}
  {arXiv:1703.03797 [hep-th]} \BibitemShut {NoStop}%
%%CITATION = ARXIV:1703.03797;%%
\bibitem [{\citenamefont {Crisostomi}\ \emph {et~al.}(2019)\citenamefont
  {Crisostomi}, \citenamefont {Koyama}, \citenamefont {Langlois}, \citenamefont
  {Noui},\ and\ \citenamefont {Steer}}]{Crisostomi:2018bsp}%
  \BibitemOpen
  \bibfield  {author} {\bibinfo {author} {\bibfnamefont {M.}~\bibnamefont
  {Crisostomi}}, \bibinfo {author} {\bibfnamefont {K.}~\bibnamefont {Koyama}},
  \bibinfo {author} {\bibfnamefont {D.}~\bibnamefont {Langlois}}, \bibinfo
  {author} {\bibfnamefont {K.}~\bibnamefont {Noui}}, \ and\ \bibinfo {author}
  {\bibfnamefont {D.~A.}\ \bibnamefont {Steer}},\ }\href {\doibase
  10.1088/1475-7516/2019/01/030} {\bibfield  {journal} {\bibinfo  {journal}
  {JCAP}\ }\textbf {\bibinfo {volume} {1901}},\ \bibinfo {pages} {030}
  (\bibinfo {year} {2019})},\ \Eprint {http://arxiv.org/abs/1810.12070}
  {arXiv:1810.12070 [hep-th]} \BibitemShut {NoStop}%
%%CITATION = ARXIV:1810.12070;%%
\bibitem [{\citenamefont {Gao}\ and\ \citenamefont {Yao}(2019)}]{Gao:2018znj}%
  \BibitemOpen
  \bibfield  {author} {\bibinfo {author} {\bibfnamefont {X.}~\bibnamefont
  {Gao}}\ and\ \bibinfo {author} {\bibfnamefont {Z.-B.}\ \bibnamefont {Yao}},\
  }\href {\doibase 10.1088/1475-7516/2019/05/024} {\bibfield  {journal}
  {\bibinfo  {journal} {JCAP}\ }\textbf {\bibinfo {volume} {1905}},\ \bibinfo
  {pages} {024} (\bibinfo {year} {2019})},\ \Eprint
  {http://arxiv.org/abs/1806.02811} {arXiv:1806.02811 [gr-qc]} \BibitemShut
  {NoStop}%
%%CITATION = ARXIV:1806.02811;%%
\bibitem [{\citenamefont {Gao}\ \emph {et~al.}(2019)\citenamefont {Gao},
  \citenamefont {Kang},\ and\ \citenamefont {Yao}}]{Gao:2019lpz}%
  \BibitemOpen
  \bibfield  {author} {\bibinfo {author} {\bibfnamefont {X.}~\bibnamefont
  {Gao}}, \bibinfo {author} {\bibfnamefont {C.}~\bibnamefont {Kang}}, \ and\
  \bibinfo {author} {\bibfnamefont {Z.-B.}\ \bibnamefont {Yao}},\ }\href
  {\doibase 10.1103/PhysRevD.99.104015} {\bibfield  {journal} {\bibinfo
  {journal} {Phys. Rev.}\ }\textbf {\bibinfo {volume} {D99}},\ \bibinfo {pages}
  {104015} (\bibinfo {year} {2019})},\ \Eprint
  {http://arxiv.org/abs/1902.07702} {arXiv:1902.07702 [gr-qc]} \BibitemShut
  {NoStop}%
%%CITATION = ARXIV:1902.07702;%%
\bibitem [{\citenamefont {Gao}(2014)}]{Gao:2014soa}%
  \BibitemOpen
  \bibfield  {author} {\bibinfo {author} {\bibfnamefont {X.}~\bibnamefont
  {Gao}},\ }\href {\doibase 10.1103/PhysRevD.90.081501} {\bibfield  {journal}
  {\bibinfo  {journal} {Phys. Rev.}\ }\textbf {\bibinfo {volume} {D90}},\
  \bibinfo {pages} {081501} (\bibinfo {year} {2014})},\ \Eprint
  {http://arxiv.org/abs/1406.0822} {arXiv:1406.0822 [gr-qc]} \BibitemShut
  {NoStop}%
%%CITATION = ARXIV:1406.0822;%%
\bibitem [{\citenamefont {Motohashi}\ and\ \citenamefont
  {Mukohyama}(2020)}]{Motohashi:2019ymr}%
  \BibitemOpen
  \bibfield  {author} {\bibinfo {author} {\bibfnamefont {H.}~\bibnamefont
  {Motohashi}}\ and\ \bibinfo {author} {\bibfnamefont {S.}~\bibnamefont
  {Mukohyama}},\ }\href {\doibase 10.1088/1475-7516/2020/01/030} {\bibfield
  {journal} {\bibinfo  {journal} {JCAP}\ }\textbf {\bibinfo {volume} {2001}},\
  \bibinfo {pages} {030} (\bibinfo {year} {2020})},\ \Eprint
  {http://arxiv.org/abs/1912.00378} {arXiv:1912.00378 [gr-qc]} \BibitemShut
  {NoStop}%
%%CITATION = ARXIV:1912.00378;%%
\bibitem [{\citenamefont {Motohashi}\ and\ \citenamefont
  {Hu}(2015)}]{Motohashi:2015hpa}%
  \BibitemOpen
  \bibfield  {author} {\bibinfo {author} {\bibfnamefont {H.}~\bibnamefont
  {Motohashi}}\ and\ \bibinfo {author} {\bibfnamefont {W.}~\bibnamefont {Hu}},\
  }\href {\doibase 10.1103/PhysRevD.92.043501} {\bibfield  {journal} {\bibinfo
  {journal} {Phys. Rev.}\ }\textbf {\bibinfo {volume} {D92}},\ \bibinfo {pages}
  {043501} (\bibinfo {year} {2015})},\ \Eprint
  {http://arxiv.org/abs/1503.04810} {arXiv:1503.04810 [astro-ph.CO]}
  \BibitemShut {NoStop}%
%%CITATION = ARXIV:1503.04810;%%
\bibitem [{\citenamefont {Gourgoulhon}(2007)}]{Gourgoulhon:2007ue}%
  \BibitemOpen
  \bibfield  {author} {\bibinfo {author} {\bibfnamefont {E.}~\bibnamefont
  {Gourgoulhon}},\ }\href@noop {} {\  (\bibinfo {year} {2007})},\ \Eprint
  {http://arxiv.org/abs/gr-qc/0703035} {arXiv:gr-qc/0703035 [GR-QC]}
  \BibitemShut {NoStop}%
%%CITATION = GR-QC/0703035;%%
\bibitem [{\citenamefont {Passaglia}\ and\ \citenamefont
  {Hu}(2018)}]{Passaglia:2018afq}%
  \BibitemOpen
  \bibfield  {author} {\bibinfo {author} {\bibfnamefont {S.}~\bibnamefont
  {Passaglia}}\ and\ \bibinfo {author} {\bibfnamefont {W.}~\bibnamefont {Hu}},\
  }\href {\doibase 10.1103/PhysRevD.98.023526} {\bibfield  {journal} {\bibinfo
  {journal} {Phys. Rev.}\ }\textbf {\bibinfo {volume} {D98}},\ \bibinfo {pages}
  {023526} (\bibinfo {year} {2018})},\ \Eprint
  {http://arxiv.org/abs/1804.07741} {arXiv:1804.07741 [astro-ph.CO]}
  \BibitemShut {NoStop}%
%%CITATION = ARXIV:1804.07741;%%
\bibitem [{\citenamefont {Motloch}\ \emph {et~al.}(2015)\citenamefont
  {Motloch}, \citenamefont {Hu}, \citenamefont {Joyce},\ and\ \citenamefont
  {Motohashi}}]{Motloch:2015gta}%
  \BibitemOpen
  \bibfield  {author} {\bibinfo {author} {\bibfnamefont {P.}~\bibnamefont
  {Motloch}}, \bibinfo {author} {\bibfnamefont {W.}~\bibnamefont {Hu}},
  \bibinfo {author} {\bibfnamefont {A.}~\bibnamefont {Joyce}}, \ and\ \bibinfo
  {author} {\bibfnamefont {H.}~\bibnamefont {Motohashi}},\ }\href {\doibase
  10.1103/PhysRevD.92.044024} {\bibfield  {journal} {\bibinfo  {journal} {Phys.
  Rev.}\ }\textbf {\bibinfo {volume} {D92}},\ \bibinfo {pages} {044024}
  (\bibinfo {year} {2015})},\ \Eprint {http://arxiv.org/abs/1505.03518}
  {arXiv:1505.03518 [hep-th]} \BibitemShut {NoStop}%
%%CITATION = ARXIV:1505.03518;%%
\bibitem [{\citenamefont {Motloch}\ \emph {et~al.}(2016)\citenamefont
  {Motloch}, \citenamefont {Hu},\ and\ \citenamefont
  {Motohashi}}]{Motloch:2016msa}%
  \BibitemOpen
  \bibfield  {author} {\bibinfo {author} {\bibfnamefont {P.}~\bibnamefont
  {Motloch}}, \bibinfo {author} {\bibfnamefont {W.}~\bibnamefont {Hu}}, \ and\
  \bibinfo {author} {\bibfnamefont {H.}~\bibnamefont {Motohashi}},\ }\href
  {\doibase 10.1103/PhysRevD.93.104026} {\bibfield  {journal} {\bibinfo
  {journal} {Phys. Rev.}\ }\textbf {\bibinfo {volume} {D93}},\ \bibinfo {pages}
  {104026} (\bibinfo {year} {2016})},\ \Eprint
  {http://arxiv.org/abs/1603.03423} {arXiv:1603.03423 [hep-th]} \BibitemShut
  {NoStop}%
%%CITATION = ARXIV:1603.03423;%%
\bibitem [{\citenamefont {Hu}\ and\ \citenamefont {Joyce}(2017)}]{Hu:2016wfa}%
  \BibitemOpen
  \bibfield  {author} {\bibinfo {author} {\bibfnamefont {W.}~\bibnamefont
  {Hu}}\ and\ \bibinfo {author} {\bibfnamefont {A.}~\bibnamefont {Joyce}},\
  }\href {\doibase 10.1103/PhysRevD.95.043529} {\bibfield  {journal} {\bibinfo
  {journal} {Phys. Rev.}\ }\textbf {\bibinfo {volume} {D95}},\ \bibinfo {pages}
  {043529} (\bibinfo {year} {2017})},\ \Eprint
  {http://arxiv.org/abs/1612.02454} {arXiv:1612.02454 [astro-ph.CO]}
  \BibitemShut {NoStop}%
%%CITATION = ARXIV:1612.02454;%%
\bibitem [{\citenamefont {Ijjas}(2018)}]{Ijjas:2017pei}%
  \BibitemOpen
  \bibfield  {author} {\bibinfo {author} {\bibfnamefont {A.}~\bibnamefont
  {Ijjas}},\ }\href {\doibase 10.1088/1475-7516/2018/02/007} {\bibfield
  {journal} {\bibinfo  {journal} {JCAP}\ }\textbf {\bibinfo {volume} {1802}},\
  \bibinfo {pages} {007} (\bibinfo {year} {2018})},\ \Eprint
  {http://arxiv.org/abs/1710.05990} {arXiv:1710.05990 [gr-qc]} \BibitemShut
  {NoStop}%
%%CITATION = ARXIV:1710.05990;%%
\bibitem [{\citenamefont {Dobre}\ \emph {et~al.}(2018)\citenamefont {Dobre},
  \citenamefont {Frolov}, \citenamefont {Ghersi}, \citenamefont {Ramazanov},\
  and\ \citenamefont {Vikman}}]{Dobre:2017pnt}%
  \BibitemOpen
  \bibfield  {author} {\bibinfo {author} {\bibfnamefont {D.~A.}\ \bibnamefont
  {Dobre}}, \bibinfo {author} {\bibfnamefont {A.~V.}\ \bibnamefont {Frolov}},
  \bibinfo {author} {\bibfnamefont {J.~T.~G.}\ \bibnamefont {Ghersi}}, \bibinfo
  {author} {\bibfnamefont {S.}~\bibnamefont {Ramazanov}}, \ and\ \bibinfo
  {author} {\bibfnamefont {A.}~\bibnamefont {Vikman}},\ }\href {\doibase
  10.1088/1475-7516/2018/03/020} {\bibfield  {journal} {\bibinfo  {journal}
  {JCAP}\ }\textbf {\bibinfo {volume} {03}},\ \bibinfo {pages} {020} (\bibinfo
  {year} {2018})},\ \Eprint {http://arxiv.org/abs/1712.10272} {arXiv:1712.10272
  [gr-qc]} \BibitemShut {NoStop}%
\bibitem [{\citenamefont {Lagos}\ \emph {et~al.}(2019)\citenamefont {Lagos},
  \citenamefont {Lin},\ and\ \citenamefont {Hu}}]{Lagos:2019rfc}%
  \BibitemOpen
  \bibfield  {author} {\bibinfo {author} {\bibfnamefont {M.}~\bibnamefont
  {Lagos}}, \bibinfo {author} {\bibfnamefont {M.-X.}\ \bibnamefont {Lin}}, \
  and\ \bibinfo {author} {\bibfnamefont {W.}~\bibnamefont {Hu}},\ }\href
  {\doibase 10.1103/PhysRevD.100.123507} {\bibfield  {journal} {\bibinfo
  {journal} {Phys. Rev.}\ }\textbf {\bibinfo {volume} {D100}},\ \bibinfo
  {pages} {123507} (\bibinfo {year} {2019})},\ \Eprint
  {http://arxiv.org/abs/1908.08785} {arXiv:1908.08785 [gr-qc]} \BibitemShut
  {NoStop}%
%%CITATION = ARXIV:1908.08785;%%
\bibitem [{\citenamefont {Abbott}\ \emph
  {et~al.}(2017{\natexlab{a}})\citenamefont {Abbott} \emph
  {et~al.}}]{TheLIGOScientific:2017qsa}%
  \BibitemOpen
  \bibfield  {author} {\bibinfo {author} {\bibfnamefont {B.~P.}\ \bibnamefont
  {Abbott}} \emph {et~al.} (\bibinfo {collaboration} {LIGO Scientific,
  Virgo}),\ }\href {\doibase 10.1103/PhysRevLett.119.161101} {\bibfield
  {journal} {\bibinfo  {journal} {Phys. Rev. Lett.}\ }\textbf {\bibinfo
  {volume} {119}},\ \bibinfo {pages} {161101} (\bibinfo {year}
  {2017}{\natexlab{a}})},\ \Eprint {http://arxiv.org/abs/1710.05832}
  {arXiv:1710.05832 [gr-qc]} \BibitemShut {NoStop}%
%%CITATION = ARXIV:1710.05832;%%
\bibitem [{\citenamefont {Abbott}\ \emph
  {et~al.}(2017{\natexlab{b}})\citenamefont {Abbott} \emph
  {et~al.}}]{Monitor:2017mdv}%
  \BibitemOpen
  \bibfield  {author} {\bibinfo {author} {\bibfnamefont {B.~P.}\ \bibnamefont
  {Abbott}} \emph {et~al.} (\bibinfo {collaboration} {LIGO Scientific, Virgo,
  Fermi-GBM, INTEGRAL}),\ }\href {\doibase 10.3847/2041-8213/aa920c} {\bibfield
   {journal} {\bibinfo  {journal} {Astrophys. J.}\ }\textbf {\bibinfo {volume}
  {848}},\ \bibinfo {pages} {L13} (\bibinfo {year} {2017}{\natexlab{b}})},\
  \Eprint {http://arxiv.org/abs/1710.05834} {arXiv:1710.05834 [astro-ph.HE]}
  \BibitemShut {NoStop}%
%%CITATION = ARXIV:1710.05834;%%
\bibitem [{\citenamefont {Ram\'irez}\ \emph {et~al.}(2018)\citenamefont
  {Ram\'irez}, \citenamefont {Passaglia}, \citenamefont {Motohashi},
  \citenamefont {Hu},\ and\ \citenamefont {Mena}}]{Ramirez:2018dxe}%
  \BibitemOpen
  \bibfield  {author} {\bibinfo {author} {\bibfnamefont {H.}~\bibnamefont
  {Ram\'irez}}, \bibinfo {author} {\bibfnamefont {S.}~\bibnamefont
  {Passaglia}}, \bibinfo {author} {\bibfnamefont {H.}~\bibnamefont
  {Motohashi}}, \bibinfo {author} {\bibfnamefont {W.}~\bibnamefont {Hu}}, \
  and\ \bibinfo {author} {\bibfnamefont {O.}~\bibnamefont {Mena}},\ }\href
  {\doibase 10.1088/1475-7516/2018/04/039} {\bibfield  {journal} {\bibinfo
  {journal} {JCAP}\ }\textbf {\bibinfo {volume} {1804}},\ \bibinfo {pages}
  {039} (\bibinfo {year} {2018})},\ \Eprint {http://arxiv.org/abs/1802.04290}
  {arXiv:1802.04290 [astro-ph.CO]} \BibitemShut {NoStop}%
%%CITATION = ARXIV:1802.04290;%%
\bibitem [{\citenamefont {Deffayet}\ \emph {et~al.}(2010)\citenamefont
  {Deffayet}, \citenamefont {Pujolas}, \citenamefont {Sawicki},\ and\
  \citenamefont {Vikman}}]{Deffayet:2010qz}%
  \BibitemOpen
  \bibfield  {author} {\bibinfo {author} {\bibfnamefont {C.}~\bibnamefont
  {Deffayet}}, \bibinfo {author} {\bibfnamefont {O.}~\bibnamefont {Pujolas}},
  \bibinfo {author} {\bibfnamefont {I.}~\bibnamefont {Sawicki}}, \ and\
  \bibinfo {author} {\bibfnamefont {A.}~\bibnamefont {Vikman}},\ }\href
  {\doibase 10.1088/1475-7516/2010/10/026} {\bibfield  {journal} {\bibinfo
  {journal} {JCAP}\ }\textbf {\bibinfo {volume} {1010}},\ \bibinfo {pages}
  {026} (\bibinfo {year} {2010})},\ \Eprint {http://arxiv.org/abs/1008.0048}
  {arXiv:1008.0048 [hep-th]} \BibitemShut {NoStop}%
%%CITATION = ARXIV:1008.0048;%%
\end{thebibliography}%

\end{document}